\documentclass[final,3p]{elsarticle}

\usepackage{graphicx}
\usepackage{caption}
\usepackage{subcaption}
\usepackage{comment}
\usepackage{amssymb}
\usepackage{amsmath}
\usepackage{dirtytalk}
\usepackage{algorithm}
\usepackage{algorithmic}
\biboptions{numbers,sort&compress}
\usepackage{lineno}

\usepackage{xcolor}
\usepackage{hyperref}
\usepackage{siunitx}
\usepackage{tikz}
\usepackage{setspace}

\usepackage{pgfplots}
\usetikzlibrary{matrix}
\usetikzlibrary{plotmarks}
\pgfplotsset{compat=newest}
\usepackage{makecell}
\usepackage[export]{adjustbox}

\usepackage{amsmath}

\hypersetup{      
    colorlinks=true,                
    linkcolor= blue,                
    filecolor=black,
    citecolor=blue,
    linkbordercolor=blue
}
\graphicspath{ {./Figures/} }
\definecolor{brandeisblue}{rgb}{0.0, 0.44, 1.0}

\journal{Elsevier}
\begin{document}

\begin{frontmatter}


\title{An eXtended Finite Element Method Implementation in COMSOL Multiphysics:  Thermo-Hydro-Mechanical Modeling of Fluid Flow in Discontinuous Porous Media}




\author[civilUNSW]{Ahmad Jafari}
\author[civilUNSW]{Mohammad Vahab*}
\cortext[cor]{Corresponding author:}
\ead{m.vahab@unsw.edu.au}
\author[civilShiraz]{Pooyan Broumand}
\author[civilUNSW]{Nasser Khalili}

\address[civilUNSW]{School of Civil and Environmental Engineering, University of New South Wales, Sydney 2052, Australia}
\address[civilShiraz]{Department of Civil and Environmental Engineering,  Shiraz University, Shiraz, Iran}

\begin{abstract}
This paper presents the implementation of the eXtended Finite Element Method (XFEM) in the general-purpose commercial software package COMSOL Multiphysics for multi-field thermo-hydro-mechanical problems in discontinuous porous media. To this end, an exclusive enrichment strategy is proposed in compliance with the COMSOL modeling structure. COMSOL modules and physics interfaces are adopted to take account of the relevant physical processes involved in thermo-hydro-mechanical coupling analysis, namely: the mechanical deformation, fluid flow in porous media and heat transfer. Essential changes are made to the internal variables of the physics interfaces to ensure consistency in the evaluation of enriched solution fields. The model preprocessing, level-set updates, coupling of the relevant physics and postprocessing procedures are performed adopting a coherent utilization of the COMSOL’s built-in features along with the COMSOL's LiveLink for MATLAB functions. The implementation process, remedies for the treatment of the enriched zones, XFEM framework setup, multiphysics coupling, numerical integration and numerical solution strategy are described in detail. The capabilities and performance of the proposed approach are investigated by examining several multi-field thermo-hydro-mechanical simulations involving single/multiple discontinuities in 2D/3D porous rock settings.

\end{abstract}


\end{frontmatter}

\section{Introduction}
\label{S:1 (introduction)}

The thermo-hydro-mechanical (THM) coupled processes due to fluid flow in deformable porous media, subject to natural discontinuities, has been a crucial area of interest for the prediction of the physical response in many engineering problems in geotechnics, mining, petroleum engineering, water resources, reservoir engineering and environmental engineering, to name a few \cite{khalili2001elasto,khalili2003fully, gelet2012thermo, bower1997numerical, watanabe2010uncertainty,sarris2012modeling}. Over the past decades, concurrent experimental \cite{zhang2020acoustic,zhou2011suggested}, analytical \cite{terzaghi1996soil} and numerical \cite{zienkiewicz1999computational, coussy2004poromechanics, sarris2011influence, tamizdoust2020fully, iranmanesh2018extrinsically} efforts were undertaken to develop predictive tools for the analysis of the role of discontinuities on the THM processes in deformable porous media \cite{gawin2006hygro,khalili2003fully}. With excellent inherent flexibility in tackling all types of discontinuities, computational superiority, and other algorithmic advantages (e.g., circumventing the need for remeshing, data transfer, and mesh refinement for high gradients), extended finite element method (XFEM) \cite{Belytschko1999XFEM, moes1999finite, dolbow2000discontinuous, sukumar2000extended} has emerged as one of the most versatile tools for the study of discontinuities in deformable porous rock media. In this respect, XFEM has been extensively adopted for the extension of hydro-mechanical \cite{rethore2007two,rethore2008two,khoei2011numerical,khoei2014mesh,salimzadeh2015three,haddad2016xfem,khoei2018enriched, vahab2019x, cruz2019xfem,parchei2020dynamic, mikaeili2018xfem,jafari2021fully,jin2020fluid} and thermo-hydro-mechanical \cite{khoei2012thermo, li2016fully, khoei2018application, nguyen2019identification,zeng2020extended,khoei2020thermo} frameworks to the analysis of fluid flow within partially/fully saturated deformable porous media in the presence of discontinuities.

Nevertheless, the application of XFEM to real-life engineering problems has been relatively limited, particularly with respect to the analysis of multiphysics processes in porous media. This has primarily been due to the fact that the majority of contributions in this context have been due to in-house codes \cite{flemisch2018benchmarks, keilegavlen2021porepy}, with limited availability and capacity for handling complex geometries, large-scale domain and 3D analysis. To alleviate this shortcoming, implementations of XFEM in general-purpose FE packages, such as ABAQUS, has been proposed through development of an elaborate host of user-defined subroutines with applications in linear elastic fracture mechanics (LEFM) by Giner et al. \cite{giner2009abaqus}, cohesive fractures by Xu and Yuan \cite{xu2009damage} and Haddad and Sepehrnoori \cite{haddad2016xfem}, and contact mechanics by Ooi et al. \cite{ooi2018investigating}. Such implementations of XFEM, despite featuring efficient built-in solvers and advanced meshing tools in 2D/3D settings, have been limited to a limited range of physics, hindering the generalized applicability of XFEM to the study of multi-field problems.

COMSOL Multiphysics is considered as one of the pioneering software packages which enables efficient handling of intricate coupling systems with enhanced efficiency and reliability \cite{pepper2017finite,hokr2018situ,michalec2021fully}. It facilitates both fully-coupled (monolithic) and segregated (staggered) solution strategies \cite{segura2008coupled}, which can be user-customized depending on the interplaying physics, accessibility of computational resources, and numerical stability concerns. To date, COMSOL has been employed as a versatile platform for the phase-field implementation of a coupled hydro-mechanical analysis of porous media (e.g., see Zhou et al.  \cite{zhou2018phase,zhou2019phase}). However, no such extension of this powerful platform has been made to include XFEM within a multi-physics platform for multi-field practical solutions. 

In this work, underpinned by our recent developments on the implementation of XFEM for solid mechanics \cite{jafari2021XFEM}, and by taking advantage of the exceptional flexibility of this software in dealing with any arbitrary coupling processes, we present a thermo-hydro-mechanical framework for the XFEM analysis of discontinuities in deformable porous media. In this context, An exclusive XFEM framework for the study of THM coupling processes, compatible with the structure of COMSOL is introduced. The THM coupling process is carried out by exploiting the generic “Solid Mechanics”, “Darcy’s Law” and “Heat Transfer in Porous Media” physics interfaces in a conjugate manner to take account of the standard and enriched terms of the solution field. The discontinuity interfaces are tracked by means of level set functions introduced via external MATLAB functions, which accommodate the evolution of cracks prior to (i.e., at the pre-processing stage) and during the analysis. Moreover, by using internal functions and variables of the software, stress intensity factor calculation as well as numerical contact analysis are performed within the Multiphysics theme. The proposed framework enables robust and efficient tackling of coupling systems with enhanced reliability in 2D/3D settings. The extended framework, in turn, facilitates the inclusion of further related physics as required.

The paper is organized as follows: In section \ref{S:2 (Formulation)}, the XFEM formulation of thermo-hydro-mechanical coupling in discontinuous deformable porous media is briefly described in conjunction with the weak forms and discretization. In section \ref{S:3 COMSOL impl}, the algorithms employed for the implementation of XFEM in COMSOL are presented, which involves a detailed discussion on the essential steps related to pre-processing of the interfaces and set-up of modules. Section \ref{S:4 (results)} is devoted to the numerical simulations which explore the performance of the extended framework in dealing with a range of benchmark problems in the literature. The concluding remarks are presented in section \ref{S:5 (Conclusions)}. The models developed are available from \href{https://github.com/ahmadjafari93/xfem-comsol.git}{https://github.com/ahmadjafari93/xfem-comsol.git}.

\section{XFEM formulation}
\label{S:2 (Formulation)}
In the context of the XFEM, the \textit{weak/strong} interfaces such as cracks, voids, and material heterogeneities in the domain are introduced independently from the background mesh by taking advantage of the partition of unity concept \cite{PUM1997}. Such description is achieved by enriching the standard finite element discretization with special enrichment functions corresponding to the desired solution fields \cite{khoei2014extended}. In the case of thermo-hydro-mechanical modeling of fluid flow in porous media, the fully coupled model is constructed by means of three independent yet interacting sub-models \cite{khoei2012thermo}; a poro-elastic model for the solid skeleton, a Darcy model for the fluid flow through the pore devoid, and a heat transfer model for the pore-solid mixture. In the following sections, the strong forms of the governing equations of each of the processes involved are presented in detail. Next, the corresponding weak forms and finite element discretization are obtained in compliance with the structure of the software \cite{borja2008assumed,jafari2021XFEM}.

\subsection{Governing equations}
\label{S:2-1 governing eqs}
\begin{figure}[!t]
\centering\includegraphics[width=0.75\linewidth]{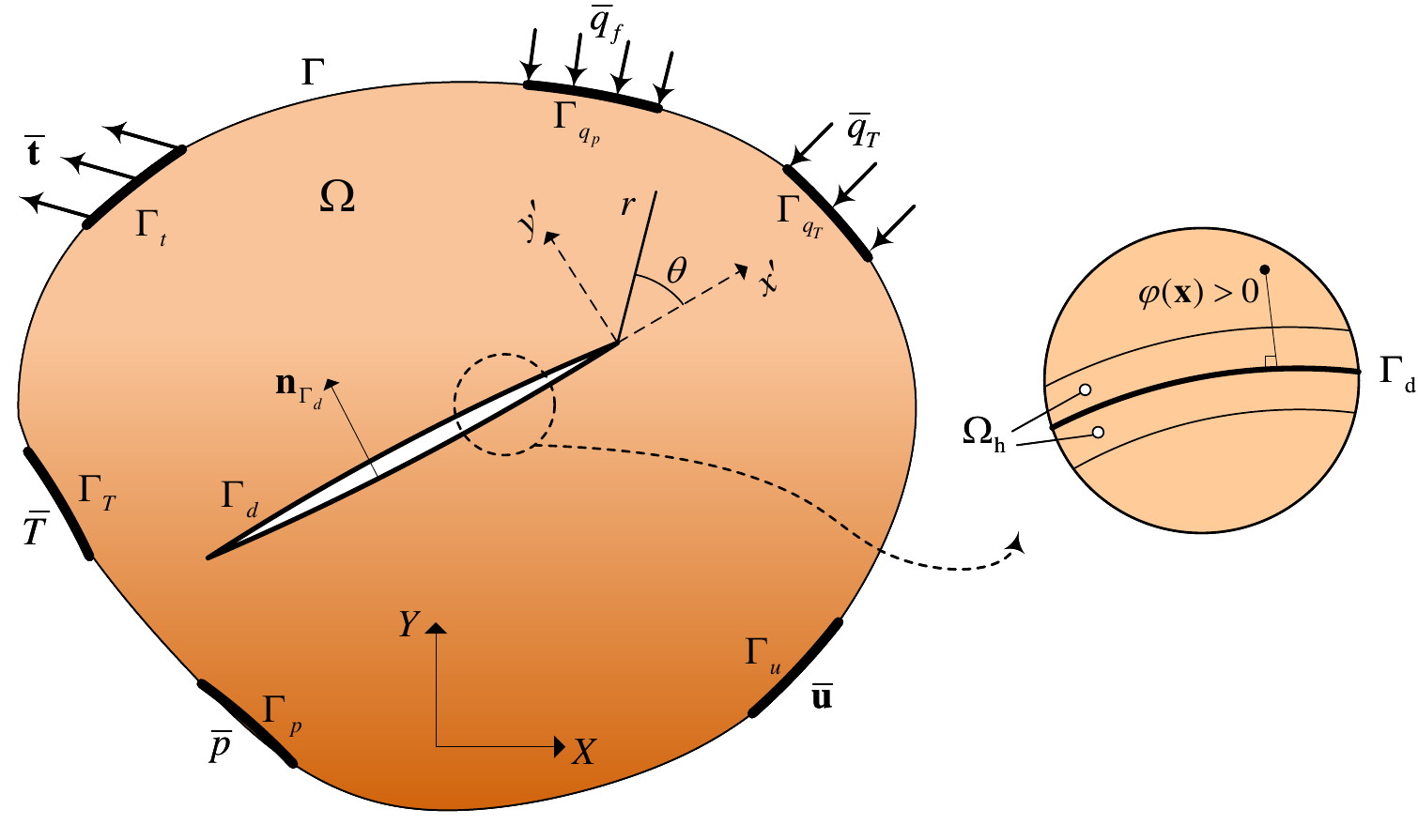}
\caption{Schematic representation of the porous problem domain, boundary conditions and discontinuity zone for the corresponding thermo-hydro-mechanical analysis.}
\label{fig: potato}
\end{figure}

The thermo-hydro-mechanical model is based on Biot's theory of the pore fluid-solid mixture \cite{biot1956general} in conjunction with the heat transfer equations. The linear momentum balance equation for the discontinuous porous domain $\Omega $ which is bounded by $\Gamma$ (Fig. \ref{fig: potato}) can be expressed as
\begin{equation}
\label{eq:equilibrium}
\nabla  \cdot {\boldsymbol{\sigma}} + \rho {\bf{b}} = 0
\end{equation}
where ${\boldsymbol{\sigma}}$ is the total Cauchy's stress, ${\bf{b}}$ denotes the body force vector and $\rho $ is weighted average density of the mixture. Based on Biot's theory, the effective stress correlates with the total stress tensor as ${\boldsymbol{\sigma }'} = {\boldsymbol{\sigma}} - \alpha p{\bf{I}}$, where ${\boldsymbol{\sigma}'}$ is the effective stress, $p$ is the pore fluid pressure and ${\bf{I}}$ is the identity tensor. For an isotropic porous medium, the scalar Biot's coefficient $\alpha$ is defined as $\alpha  = 1 - \frac{{{K_{\rm{t}}}}}{{{K_{\rm{s}}}}}$, in which ${{K_{\rm{t}}}}$ and ${{K_{\rm{s}}}}$ are respectively the bulk moduli of the porous medium and solid particles. Moreover, $\rho  = (1 - n){\rho _{\rm{s}}} + n{\rho _{\rm{f}}}$, where ${\rho _{\rm{s}}}$ and ${\rho _{\rm{f}}}$ respectively represent the density of the solid and fluid phases, and $n$ is the porosity of the whole mixture. 

The thermally induced strain is expressed as 
\begin{equation}
\label{eq: thermal strain}
{{\boldsymbol{\varepsilon }}_T} =  {\boldsymbol{\alpha} _T}(T - {T_{{\rm{0}}}})
\end{equation}
where $T$ and ${T_{{\rm{0}}}}$  are the current and reference temperatures of the porous medium. $\boldsymbol{\alpha }_T$ is the thermal expansivity tensor, which is defined for an isotropic medium as $\boldsymbol{\alpha} _T = {\beta _{\rm{s}}}{\bf{I}}$. Here ${{\beta _{\rm{s}}}}$ is the volumetric thermal expansion coefficient of the porous domain, which is identical to thermal expansion coefficient of the solid constituent \cite{khalili2010skeletal}. It is postulated that the solid-fluid mixture is locally in thermal equilibrium since the fluid velocity inside the pores is relatively insignificant \cite{salimzadeh2018three}. Therefore, the constitutive equations for the effective stress can be expressed as \cite{khalili2003fully}:
\begin{equation}
\label{eq: stress-strain}
{\boldsymbol{\sigma}}' = {\bf{D}}{\boldsymbol{\varepsilon }} - \left[ {\beta _{\rm{s}}}{K_{\rm{t}}}(T - {T_\text{0}})\right] {\bf{I}} 
\end{equation}
where ${\bf{D}}$ is the Hooke’s elasticity tensor and ${\boldsymbol{\varepsilon }} = \frac{1}{2}\left( {\nabla {\bf{u}} + {\nabla ^{\rm{T}}}{\bf{u}}} \right)$ is the total infinitesimal strain tensor, with ${\bf{u}}$ being the displacement vector. In this study, internal discontinuities are considered as impermeable boundaries with zero leak-off. In this fashion, the continuity equation of the fluid phase in the porous matrix can be obtained by employing the averaging theory in the porous media as follows \cite{khalili2001elasto}:
\begin{equation}
\label{eq: continuity}
\alpha \nabla  \cdot {\dot{\bf{u}}} + \nabla  \cdot {\dot{\bf{w}}_{\rm{f}}} + \frac{1}{{{Q_{\rm{t}}}}}\dot p - {\beta _{\rm{t}}}\dot T = 0
\end{equation}
where $\frac{1}{{{Q_{\rm{t}}}}} = \left( {\frac{{\alpha  - n}}{{{K_{\rm{s}}}}} + \frac{n}{{{K_{\rm{f}}}}}} \right)$ is the compressibility coefficient of the bulk medium. Moreover, ${\beta _{\rm{t}}} = {(\alpha  - n){\beta _{\rm{s}}} + n{\beta _{\rm{f}}}} $, where ${\beta_{\rm{f}}}$ denotes the fluid coefficients of thermal expansion. ${\dot{\bf{u}}}$ is the solid phase velocity, and ${\dot{\bf{w}}_{\rm{f}}}$ is the relative velocity of the fluid phase with respect to the solid skeleton that is expressed by means of Darcy’s law as
\begin{equation}
\label{eq: Darcy}
{\dot{\bf{ w}}_{\rm{f}}} = \frac{{{k_{\rm{f}}}}}{{{\mu _{\rm{f}}}}}\left( { - \nabla p + {\rho _{\rm{f}}}{\bf{b}}} \right)
\end{equation}
in which ${{k_{\rm{f}}}}$ and ${{\mu _{\rm{f}}}}$ are the intrinsic permeability of the domain and fluid viscosity, respectively. Combining Eqs. \ref{eq: continuity} and \ref{eq: Darcy}, the flow-continuity equation of the pore fluid can be recast into
\begin{equation}
\label{eq: flow-continuity}
\alpha \nabla  \cdot {\dot{\bf{u}}} + \nabla  \cdot \left( {\frac{{{k_{\rm{f}}}}}{{{\mu _{\rm{f}}}}}\left( { - \nabla p + {\rho _{\rm{f}}}{\bf{b}}} \right)} \right) + \frac{1}{{{Q_{\rm{t}}}}}\dot p - {\beta _{\rm{t}}}\dot T = 0
\end{equation}

The heat transfer relation is deduced by combining Fourier’s law and the conservation of energy equation for the porous media. In resemblance to the fluid flow formulation, the thermal energy balance equation is derived using the averaging theory for the energy balance of both solid and fluid phases, as follows \cite{lewis1998finite}
\begin{equation}
\label{eq: energy balance}
{\left( {\rho C} \right)_{\text{eff}}}\dot T + {\rho _{\rm{f}}}{C_{\rm{f}}} \left( {\frac{{{k_{\rm{f}}}}}{{{\mu _{\rm{f}}}}}\left( { - \nabla p + {\rho _{\rm{f}}}{\bf{b}}} \right)} \right) \cdot \nabla T - \nabla  \cdot \left( {{\lambda _{\text{eff}}}\nabla T} \right) = 0
\end{equation}
where ${ C_{\text{eff}}}$ and ${\lambda _\text{eff}}$ are the effective specific heat capacity and thermal conductivity, respectively, defined as
\begin{equation}
\label{eq: capacity conductivity}
\begin{array}{l}
{\left( {\rho C} \right)_{\text{eff}}} = (1 - n){\rho _{\rm{s}}}{C_{\rm{s}}} + n{\rho _{\rm{f}}}{C_{\rm{f}}}\\
{\lambda _{\text{eff}}} = (1 - n){\lambda _{\rm{s}}} + n{\lambda _{\rm{f}}}
\end{array}
\end{equation}
In Eq. \ref{eq: capacity conductivity}, ${C_{\rm{s}}}$, ${C_{\rm{f}}}$, ${\lambda _{\rm{s}}}$ and ${\lambda _{\rm{f}}}$ are the specific heat capacities and thermal heat conductivities of the solid and fluid phases, respectively. It is worth noting that heat generation due to phase compression is neglected \cite{khalili2001elasto}.

The above thermo-hydro-mechanical governing equations are subjected to the following Dirichlet boundary conditions on the external boundaries:
\begin{equation}
\label{eq: essential BC external}
\begin{array}{*{20}{c}}
{{\bf{u = \bar u}}}&{{\rm{on}}}&{{\Gamma _u}}\\
{p = \bar p}&{{\rm{on}}}&{{\Gamma _p}}\\
{T = \bar T}&{{\rm{on}}}&{{\Gamma _T}}
\end{array}
\end{equation}
where ${{\bf{\bar u}}}$, ${\bar p}$ and ${\bar T}$ are the prescribed displacement, pressure and temperature applied on the associated external boundaries ${{\Gamma _u}}$, ${{\Gamma _p}}$ and ${{\Gamma _T}}$, respectively. The Neumann boundary conditions on the contrary are described as
\begin{equation}
\label{eq: natural BC external}
\begin{array}{*{20}{c}}
{{\boldsymbol{\sigma}} \cdot {{\bf{n}}_{\Gamma }=\bar{\bf  t}}}&{{\rm{on}}}&{{\Gamma _{\rm{t}}}}\\
{{{\dot{\bf{w}}}} \cdot {{\bf{n}}_{\Gamma}} = {{\bar q}_{\rm{f}}}}&{{\rm{on}}}&{{\Gamma _{{q_p}}}}\\
{ - \left( {{\lambda _{\text{eff}}}\nabla T} \right) \cdot {{\bf{n}}_{\Gamma}} = {{\bar q}_T}}&{{\rm{on}}}&{{\Gamma _{{q_T}}}}
\end{array}
\end{equation}
where ${{{\bf{n}}_{\Gamma}}}$ is the normal unit vector of the external boundary $\Gamma $. In addition, ${{\bar{\bf t}}}$, ${{{\bar q}_{\rm{f}}}}$ and ${{{\bar q}_T}}$ are respectively the applied traction and, fluid and heat fluxes along the corresponding external boundaries ${{\Gamma _{\rm{t}}}}$, ${{\Gamma _{{q_p}}}}$ and ${{\Gamma _{{q_T}}}}$. Note the external boundaries are defined such that ${\Gamma _u} \cup {\Gamma _{\rm{t}}} = {\Gamma _p} \cup {\Gamma _{{q_p}}} = {\Gamma _T} \cup {\Gamma _{{q_T}}}=\Gamma$ and ${\Gamma _u} \cap {\Gamma _{\rm{t}}} = {\Gamma _p} \cap {\Gamma _{{q_p}}} = {\Gamma _T} \cap {\Gamma _{q_T}}=\emptyset $ hold.

The boundary conditions associated with the impermeable adiabatic internal discontinuities of the domain subject to fluid flow, thermal flux and contact constraints are expressed as
\begin{equation}
\label{eq: natural BC internal}
\begin{array}{*{20}{c}}
{{\boldsymbol{\sigma}} \cdot {{\bf{n}}_{{{\Gamma }_{\rm{d}}}}}{ = }{{\bf{t}}^{{\rm{c}}}}}&{{\rm{on}}}&{{\Gamma _{\rm{c}}} \subset {\Gamma _{\rm{d}}}}\\
{\left[\kern-0.15em\left[ {{{\dot{\bf{ w}}}_{\rm{f}}}} 
 \right]\kern-0.15em\right] \cdot {{\bf{n}}_{{{\bf{\Gamma }}_{\rm{d}}}}} = 0}&{{\rm{on}}}&{{\Gamma _{\rm{d}}}}\\
{ - \left[\kern-0.15em\left[ {{\lambda _{{\rm{eff}}}}\nabla T} 
 \right]\kern-0.15em\right] \cdot {{\bf{n}}_{{{\Gamma }_{\rm{d}}}}} = {q_T^{\rm{c}}}}&{{\rm{on}}}&{{\Gamma _{\rm{c}}} \subset {\Gamma _{\rm{d}}}}
\end{array}
\end{equation}
in which ${\Gamma_{\rm{d}}}$ and ${{\bf{n}}_{{{\Gamma}}_{\rm{d}}}}$ are the discontinuity interface and its normal unit vector, respectively. The notation $\left[\kern-0.15em\left[ \Re  
 \right]\kern-0.15em\right]$ represents the jump in the field $\Re \in \text{\{} {\bf{u}},{{{{\bf{ w}}}_{\rm{f}}}}, T \text{\}}$ across the discontinuity faces. Moreover, ${{\bf{t}}^{\rm{c}}}$ and $q_T^{\rm{c}}$ represent contact traction vector and heat flux, respectively, that are imposed on the contact boundary ${\Gamma _{\rm{c}}}$. Given the normal gap ${g_{\rm{N}}} = \left[\kern-0.15em\left[ {{u_{\rm{N}}}} 
 \right]\kern-0.15em\right]=\left[\kern-0.15em\left[ {{{{\bf{ u}}}}} 
 \right]\kern-0.15em\right]\cdot {{\bf{n}}_{{{\Gamma }_{\rm{d}}}}}$ (i.e., the normal component of the displacement jump), the Kuhn-Tucker criterion is employed to elaborate the impenetration condition across the discontinuity faces as follows 
 \begin{equation}
\label{eq: kuhn-tucker}
{g_{\rm{N}}} \ge 0,\,\,t_{\rm{N}}^{{\rm{c}}} \le 0,\,\,{g_{\rm{N}}}t_{\rm{N}}^{{\rm{c}}} = 0
\end{equation}
where $t_{\rm{N}}^{\rm{c}} = {{\bf{t}}^{\rm{c}}} \cdot {\textbf{n}_{{\Gamma _{\rm{d}}}}}$ is the normal contact traction along the discontinuity. Similarly, $q_T^{\rm{c}} = {h_{{\rm{cont}}}}\left[\kern-0.15em\left[ T 
 \right]\kern-0.15em\right]$ is the heat flux imposed on the mechanical contact surface, with ${h_{{\rm{cont}}}}$ and $\left[\kern-0.15em\left[ T 
 \right]\kern-0.15em\right]$ being the thermal conductivity coefficients and temperature jump across the contact interface, respectively \cite{khoei2018application}. Note that since the discontinuity interfaces are stipulated as impermeable boundaries, no fluid exchange and associated convective heat transfer may occur between the discontinuity and the surrounding porous medium. Furthermore, the conductive heat transfer is assumed to be negligible in the case of discontinuity opening. This is a valid assumption wherever the thermal conductivity is close to zero, e.g., heat insulation of the void/air inside the cavity, in the absence of any fluids \cite{madhusudana1986contact}.

\subsection{Weak form and XFEM discretization}
\label{S:2-2 weak forms XFEM dis}

In order to derive the weak form of the governing equations, i.e., Eqs. \ref{eq:equilibrium}, \ref{eq: flow-continuity} and \ref{eq: energy balance}, three sets of costume tailored tests functions ${\boldsymbol{\eta }}_u^{}$,  $\eta _p^{}$ and $\eta _T^{}$ are respectively adopted, which all satisfy the homogeneous Dirichlet boundary conditions. Multiplying the governing equations by the test functions and employing the Gauss-Divergence theorem yields the weak form equations as
\begin{equation}
\label{eq: weak form}
\begin{array}{l}
\int\limits_\Omega ^{} {{\nabla ^{\rm{s}}}{{\boldsymbol{\eta }}_u}:{\boldsymbol{\sigma}}{\rm{d}}\Omega } = \int\limits_\Omega ^{} {{{\boldsymbol{\eta }}_u} \cdot \rho {\bf{b}}{\rm{d}}\Omega }  + \int\limits_{{\Gamma _{\rm{t}}}}^{} {{{\boldsymbol{\eta }}_u} \cdot {\bf{\bar t}}{\rm{d}}\Gamma } \\
\int\limits_\Omega ^{} {{\eta _p}\alpha \nabla  \cdot {\dot{\bf{u}}}{\rm{d}}\Omega }  + \int\limits_\Omega ^{} {\nabla {\eta _p}\cdot \frac{{{k_{\rm{f}}}}}{{{\mu _{\rm{f}}}}}\nabla p{\rm{d}}\Omega }  + \int\limits_\Omega ^{} {{\eta _p}\frac{1}{{{Q_{\rm{t}}}}}\dot p{\rm{d}}\Omega }  - \int\limits_\Omega ^{} {{\eta _p}{\beta _{\rm{t}}}\dot T{\rm{d}}\Omega }  = \int\limits_{{\Gamma _{{q_p}}}}^{} {{\eta _p}{{\bar q}_{\rm{f}}}{\rm{d}}\Gamma }  - \int\limits_\Omega ^{} {{\nabla{\eta _p}}\cdot \frac{{{k_{\rm{f}}}}}{{{\mu _{\rm{f}}}}}{\rho _{\rm{f}}}{\bf{b}} \rm{d} \Omega } \\
\int\limits_\Omega ^{} {{\eta _T}{{(\rho C)}_{{\rm{eff}}}}\dot T{\rm{d}}\Omega }  + \int\limits_\Omega ^{} {{\eta _T}\left[ {{\rho _{\rm{f}}}{C_{\rm{f}}}\frac{{{k_{\rm{f}}}}}{{{\mu _{\rm{f}}}}}( - \nabla p + {\rho_{\rm{f}}} {\bf{b}})} \right]\cdot \nabla T{\rm{d}}\Omega }  + \int\limits_\Omega ^{} {\nabla {\eta _T}\cdot{\lambda _{{\rm{eff}}}}\nabla T{\rm{d}}\Omega }  = \int\limits_{{\Gamma _{{q_T}}}}^{} {{\eta _T}{{\bar q}_T}{\rm{d}}\Gamma } 
\end{array}
\end{equation}

As depicted in Fig. \ref{fig: potato}, the displacement, pressure and temperature fields are discontinuous across ${\Gamma _{\rm{d}}}$. Thus, the corresponding discretized XFEM fields can be expressed as
\begin{equation}
\label{eq: def u,p,T}
\begin{array}{l}
{\bf{u}}({\bf{x}}) = {{\bf{u}}^{{\rm{c}}}}({\bf{x}}) + H_{\Gamma _{d}}(\varphi (\mathbf{x})){{\bf{u}}^{{\rm{d}}}}({\bf{x}})\\
p({\bf{x}}) = {p^{{\rm{c}}}}({\bf{x}}) + H_{\Gamma _{d}}(\varphi (\mathbf{x})){p^{{\rm{d}}}}({\bf{x}})\\
T({\bf{x}}) = {T^{{\rm{c}}}}({\bf{x}}) + H_{\Gamma _{d}}(\varphi (\mathbf{x})){T^{{\rm{d}}}}({\bf{x}})
\end{array}
\end{equation}
where ${{\bf{u}}^{{\rm{c}}}}$, ${p^{{\rm{c}}}}$, ${T^{{\rm{c}}}}$ and ${{\bf{u}}^{{\rm{d}}}}$, ${p^{{\rm{d}}}}$, ${T^{{\rm{d}}}}$ are associated with the standard (continuous) and enriched (discontinuous) approximations of the displacement, pressure and temperature fields, respectively. Moreover, $H_{\Gamma _{d}}(\varphi (\mathbf{x}))=\mathbb{H}_{\Gamma _{d}}(\varphi (\mathbf{x}))-\mathbb{H}_{\Gamma _{d}}(\varphi (\mathbf{x})^{I})$ is the shifted Heaviside enrichment function that reproduces strong discontinuity across ${\Gamma _{\rm{d}}}$ \cite{liu2008contact}, in which
\begin{equation}
\label{eq:Heavisied}
\mathbb{H}_{\Gamma _{d}}(\varphi (\mathbf{x}))=\left\{\begin{matrix}
1 & \varphi (\mathbf{x})\geq 0\\ 
-1 & \varphi (\mathbf{x})< 0
\end{matrix}\right.
\end{equation}
is the Heaviside function and $\varphi (\mathbf{x})$ represents the signed distance function with respect to discontinuity interface $\Gamma _{d}$ \cite{khoei2014extended}. According to Eq. \ref{eq: def u,p,T}, the infinitesimal strain tensor as well as the pressure and temperature gradients can be approximated as
\begin{equation}
\label{eq: grad u,p,T}
\begin{array}{l}
{\boldsymbol{\varepsilon }}({\bf{x}}) = {\nabla ^{\rm{s}}}{\bf{u}}({\bf{x}}) = {\nabla ^{\rm{s}}}{{\bf{u}}^{{\rm{c}}}}({\bf{x}}) + {H_{{\Gamma _{\rm{d}}}}}(\varphi ({\bf{x}}))\nabla {{\bf{u}}^{{\rm{d}}}}({\bf{x}}) + {\delta _{{\Gamma _{\rm{d}}}}}{({{\bf{u}}^{{\rm{d}}}}({\bf{x}}) \otimes {{\bf{n}}_{{\Gamma _{\rm{d}}}}})^{\rm{s}}}\\
\nabla p({\bf{x}}) = \nabla {p^{{\rm{c}}}}({\bf{x}}) + {H_{{\Gamma _{\rm{d}}}}}(\varphi ({\bf{x}}))\nabla {p^{{\rm{d}}}}({\bf{x}}) + {\delta _{{\Gamma _{\rm{d}}}}}p({\bf{x}}){{\bf{n}}_{{\Gamma _{\rm{d}}}}}\\
\nabla T({\bf{x}}) = \nabla {T^{{\rm{c}}}}({\bf{x}}) + {H_{{\Gamma _{\rm{d}}}}}(\varphi ({\bf{x}}))\nabla {T^{{\rm{d}}}}({\bf{x}}) + {\delta _{{\Gamma _{\rm{d}}}}}T({\bf{x}}){{\bf{n}}_{{\Gamma _{\rm{d}}}}}
\end{array}
\end{equation}
where ${\nabla ^{\rm{s}}}$ and ${( \cdot )^{\rm{s}}}$ are the symmetric parts of the spatial gradient operator and tensor, respectively, and ${\delta _{{\Gamma _{\rm{d}}}}}$ is the Dirac’s delta function across $\Gamma _{\rm{d}}$.

 Similar to the trial functions, the test functions can also be decomposed into continuous and discontinuous parts as follows 
\begin{equation}
\label{eq: test functions}
\begin{array}{l}
{\boldsymbol{\eta }}_u^{} = {\boldsymbol{\eta }}_u^{{\rm{c}}} + {H_{{\Gamma _{\rm{d}}}}}(\varphi ({\bf{x}})){\boldsymbol{\eta }}_u^{{\rm{d}}}\\
\eta _p^{} = \eta _p^{{\rm{c}}} + {H_{{\Gamma _{\rm{d}}}}}(\varphi ({\bf{x}}))\eta _p^{{\rm{d}}}\\
\eta _T^{} = \eta _T^{{\rm{c}}} + {H_{{\Gamma _{\rm{d}}}}}(\varphi ({\bf{x}}))\eta _T^{{\rm{d}}}
\end{array}
\end{equation}

The presented XFEM formulation is fully compatible with the inherent structure of COMSOL Multiphysics \cite{jafari2021XFEM,borja2008assumed}. By substituting the above decomposition in Eq. \ref{eq: weak form}, the weak forms of the continuous parts of the governing equations can be obtained as
\begin{equation}
\label{eq: weak form c.}
\begin{array}{l}
\int\limits_\Omega ^{} {{\nabla ^{\rm{s}}}{{\boldsymbol{\eta }}_u^{{\rm{c}}} }:{\boldsymbol{\sigma}}{\rm{d}}\Omega } = \int\limits_\Omega ^{} {{{\boldsymbol{\eta }}_u^{{\rm{c}}} } \cdot \rho {\bf{b}}{\rm{d}}\Omega }  + \int\limits_{{\Gamma _{\rm{t}}}}^{} {{{\boldsymbol{\eta }}_u^{{\rm{c}}}} \cdot {\bf{\bar t}}{\rm{d}}\Gamma } \\
\int\limits_\Omega ^{} { {\eta _p^{{\rm{c}}}}\alpha \nabla  \cdot {\dot{\bf{u}}}{\rm{d}}\Omega }  + \int\limits_\Omega ^{} {\nabla {\eta _p^{{\rm{c}}}}\cdot \frac{{{k_{\rm{f}}}}}{{{\mu _{\rm{f}}}}}\nabla p{\rm{d}}\Omega }  + \int\limits_\Omega ^{} {{\eta _p^{{\rm{c}}}}\frac{1}{{{Q_{\rm{t}}}}}\dot p{\rm{d}}\Omega }  - \int\limits_\Omega ^{} {{\eta _p^{{\rm{c}}}}{\beta _{\rm{t}}}\dot T{\rm{d}}\Omega }  = \int\limits_{{\Gamma _{{q_p}}}}^{} {{\eta _p^{{\rm{c}}}}{{\bar q}_{\rm{f}}}{\rm{d}}\Gamma }  - \int\limits_\Omega ^{} {{\nabla}{\eta_p^{{\rm{c}}}}\cdot \frac{{{k_{\rm{f}}}}}{{{\mu _{\rm{f}}}}}{\rho _{\rm{f}}}{\bf{b}} \rm{d} \Omega } \\
\int\limits_\Omega ^{} {{\eta_T^{{\rm{c}}}}{{(\rho C)}_{{\rm{eff}}}}\dot T{\rm{d}}\Omega }  + \int\limits_\Omega ^{} {{\eta_T^{{\rm{c}}}}\left[ {{\rho _{\rm{f}}}{C_{\rm{f}}}\frac{{{k_{\rm{f}}}}}{{{\mu _{\rm{f}}}}}( - \nabla p + {\rho_{\rm{f}}} {\bf{b}})} \right]\cdot \nabla T{\rm{d}}\Omega }  + \int\limits_\Omega ^{} {\nabla {\eta _T^{{\rm{c}}}}\cdot {\lambda _{{\rm{eff}}}}\nabla T{\rm{d}}\Omega }  = \int\limits_{{\Gamma _{{q_T}}}}^{} {{\eta _T^{{\rm{c}}}}{{\bar q}_T}{\rm{d}}\Gamma } 
\end{array}
\end{equation}
and the discontinuous enriched parts as
\begin{equation}
\label{eq: weak form d.}
\begin{array}{l}
\int\limits_\Omega ^{} {\left[ {{H_{{\Gamma _{\rm{d}}}}}(\varphi ({\bf{x}})){\nabla ^{\rm{s}}}{\boldsymbol{\eta }}_u^{{\rm{d}}}} \right]:{\boldsymbol{\sigma}}{\rm{d}}\Omega } = \int\limits_\Omega ^{} {{H_{{\Gamma _{\rm{d}}}}}(\varphi ({\bf{x}})){\boldsymbol{\eta }}_u^{{\rm{d}}} \cdot \rho {\bf{b}}{\rm{d}}\Omega } + \int\limits_{{\Gamma _{\rm{t}}}}^{} {{H_{{\Gamma _{\rm{d}}}}}(\varphi ({\bf{x}})){\boldsymbol{\eta }}_u^{{\rm{d}}} \cdot {\bf{\bar t}}{\rm{d}}\Gamma }  - \int\limits_{{\Gamma _{\rm{c}}}}^{} {{\boldsymbol{\eta }}_u^{{\rm{d}}} \cdot {{\bf{t}}^{{\rm{c}}}}{\rm{d}}\Gamma } \\
\int\limits_\Omega ^{} {\left[ {{H_{{\Gamma _{\rm{d}}}}}(\varphi ({\bf{x}})) \eta _p^{{\rm{d}}}} \right]\alpha \nabla  \cdot {\dot{\bf{u}}}{\rm{d}}\Omega }  + \int\limits_\Omega ^{} {{H_{{\Gamma _{\rm{d}}}}}(\varphi ({\bf{x}}))\nabla \eta _p^{{\rm{d}}}\cdot \frac{{{k_{\rm{f}}}}}{{{\mu _{\rm{f}}}}}\nabla p{\rm{d}}\Omega }  + \int\limits_\Omega ^{} {{H_{{\Gamma _{\rm{d}}}}}(\varphi ({\bf{x}}))\eta _p^{{\rm{d}}}\frac{1}{{{Q_{\rm{t}}}}}\dot p{\rm{d}}\Omega } \\
 - \int\limits_\Omega ^{} {{H_{{\Gamma _{\rm{d}}}}}(\varphi ({\bf{x}}))\eta _p^{{\rm{d}}}{\beta _{\rm{t}}}\dot T{\rm{d}}\Omega }  = \int\limits_{{\Gamma _{{q_p}}}}^{} {{H_{{\Gamma _{\rm{d}}}}}(\varphi ({\bf{x}}))\eta _p^{{\rm{d}}}{{\bar q}_{\rm{f}}}{\rm{d}}\Gamma }  - \int\limits_\Omega ^{} {{H_{{\Gamma _{\rm{d}}}}}(\varphi ({\bf{x}})){\nabla} \eta _p^{{\rm{d}}}\cdot \frac{{{k_{\rm{f}}}}}{{{\mu _{\rm{f}}}}}{\rho _{\rm{f}}}{\bf{b}} \rm{d} \Omega } \\
\int\limits_\Omega ^{} {{H_{{\Gamma _{\rm{d}}}}}(\varphi ({\bf{x}}))\eta _T^{{\rm{d}}}{{(\rho C)}_{{\rm{eff}}}}\dot T{\rm{d}}\Omega }  + \int\limits_\Omega ^{} {{H_{{\Gamma _{\rm{d}}}}}(\varphi ({\bf{x}}))\eta _T^{{\rm{d}}}\left[ {{\rho_{\rm{f}}}{C_{\rm{f}}}\frac{{{k_{\rm{f}}}}}{{{\mu _{\rm{f}}}}}( - \nabla p + {\rho_{\rm{f}}} {\bf{b}})} \right]\cdot \nabla T{\rm{d}}\Omega } \\
 + \int\limits_\Omega ^{} {{H_{{\Gamma _{\rm{d}}}}}(\varphi ({\bf{x}}))\nabla \eta _T^{{\rm{d}}}\cdot {\lambda _{{\rm{eff}}}}\nabla T{\rm{d}}\Omega }  = \int\limits_{{\Gamma _{{q_T}}}}^{} {{H_{{\Gamma _{\rm{d}}}}}(\varphi ({\bf{x}}))\eta _T^{{\rm{d}}}{{\bar q}_T}{\rm{d}}\Gamma }  - \int\limits_{{\Gamma_{\rm{c}}}}^{} {\eta _T^{{\rm{d}}}{q_T^{\rm{c}}}{\rm{d}}\Gamma } 
\end{array}
\end{equation}
In Eq. \ref{eq: weak form d.}, the integration domain is limited to the support of $H_{\Gamma _{d}}(\varphi (\mathbf{x}))$ i.e., the enriched zone ${\Omega _{\rm{h}}}$ detected by the signed distance function.

In order to solve the integral Eqs. \ref{eq: weak form c.} and \ref{eq: weak form d.}, the spatial discretization is employed within the context of XFEM. Considering a Bubnov-Galerkin scheme, both the trial and test functions are discretized using ${\rm{C}^0}$  continuous shape functions ${N_i}({\bf{x}})$ for the primary variables $\aleph  = \left\{ {{\bf{u}},p,T} \right\}$ as
\begin{equation}
\label{eq: XFEM discretization}
\left\{ {\begin{array}{*{20}{c}}
{{\aleph ^{{\rm{c}}}}({\bf{x}}) = \sum\limits_{i \in {m_{{\rm{std}}}}} {{N_i}({\bf{x}}){{\hat X}_i}} }&{{\rm{in }}\,\Omega }\\
{{\aleph ^{{\rm{d}}}}({\bf{x}}) = \sum\limits_{i \in {m_{{\rm{enr}}}}} {{N_i}({\bf{x}}){{\tilde X}_i}} }&{{\rm{in }}\,{\Omega _{\rm{h}}}}
\end{array}} \right.
\end{equation}
where $\hat X = \left\{ {{\bf{\hat u}},\hat p,\hat T} \right\}$ and $\tilde X = \left\{ {{\bf{\tilde u}},\tilde p,\tilde T} \right\}$ are the vectors of nodal displacements, pressure and temperature for the standard and enriched parts, respectively. Moreover, ${m_{{\rm{std}}}}$ and ${m_{{\rm{enr}}}}$ denote the sets of standard and enriched nodes, respectively.
\section{Implementation in COMSOL}
\label{S:3 COMSOL impl}

\subsection{Overall framework}
\label{S:3-1 overview}

COMSOL is a general-purpose multi-field program that renders sophisticated multiphysics modeling capabilities. In this software, a combination of available built-in physics interfaces can be incorporated in conjunction with user-defined physics to investigate formidable multiphysics problems \cite{Multiphysics2019introduction, jafari2021XFEM}. The thermo-hydro-mechanical coupling analysis of fluid flow in deformable porous media, considered here, involves three distinct physics that take account of the mechanical deformation, fluid flow and heat transfer. In COMSOL, the “Solid Mechanics” physics interface facilitates the most general toolkit to perform continuum based structural analysis by solving the equations of motion endowed with suitable constitutive material behaviour. The fluid flow through the porous domain can be simulated via the “Porous Media and Subsurface Flow” module. In the case of low-velocity flows, which is commonly the case in geomechanics, the “Darcy’s Law” physics interface is employed to incorporate the flow-continuity equations within the porous medium. The “Heat Transfer” module offers heat transfer analysis in the host domain. Among the available interfaces in this module, the “Heat Transfer in Porous Media” physics interface is utilized to include the convection-diffusion equation through both solid matrix and pore fluid phases. In the conventional FE modeling approach, the discontinuity in the solution field needs to be explicitly modeled, while the mesh must conform to the internal boundaries for all physics interfaces. In contrast, based on the XFEM approach, for which the theoretical background is presented in the previous section, one can circumvent the requirement of mesh alignment with the internal interfaces. Extending the recent work of the authors on the XFEM implementation of fracturing of solid formations in COMSOL \cite{jafari2021XFEM}, three distinct COMSOL interfaces are introduced, one for each of the physics involved. The solution variables associated with these physics interfaces are recast in the set $ \left\{ {{\aleph ^{{\rm{c}}}},{\aleph ^{{\rm{d}}}}} \right\}$, where ${\aleph ^{{\rm{c}}}} = \left\{ {{{\bf{u}}^{{\rm{c}}}},{p^{{\rm{c}}}},{T^{{\rm{c}}}}} \right\}$  constitutes the continuous parts of the solution fields, and ${\aleph ^{{\rm{d}}}} = \left\{ {{{\bf{u}}^{{\rm{d}}}},{p^{{\rm{d}}}},{T^{{\rm{d}}}}} \right\}$  encompasses the discontinuous parts. For the sake of brevity, the allocated standard parts of the solution domain corresponding to the “Solid Mechanics”, “Darcy’s Law” and “Heat Transfer in Porous Media” physics interfaces are denoted by SMstd, DLstd and HTstd, respectively, and their enriched counterparts are indicated by SMenr, DLenr and HTenr, respectively, in the remainder of the paper.

Several important features such as integration tools and analytical functions are inherently incorporated within COMSOL software. However, additional \textit{external functions} are imperative to implement in order to successfully execute the various functionalities required for the XFEM simulation, such as the level-set analysis in the pre-processing stage. These additional functions/subroutines are developed in MATLAB and linked to the model through the COMSOL Live-link for MATLAB software extension \cite{jafari2021XFEM}. In the following sections, the pre-processing step, physics interfaces setup (i.e., XFEM enrichment, thermo-hydro-mechanical coupling, and thermo-mechanical contact), numerical integration scheme and the numerical solution strategy are elaborated in detail.

\subsection{Pre-processing: identification of crack geometry }
\label{S:3-2 preprocessing}

In XFEM simulations, a pre-processing phase is required prior to the analysis in order to identify the set of bisected elements by the crack interface. To this end, a conventional FEM mesh is first generated by neglecting the geometry of internal interfaces (i.e., cracks) in COMSOL. The mesh elemental and nodal information is then exported as a “.mphtxt” file to undertake the level-set analysis in a separate MATLAB script file “preprocess.m”. Accordingly, a geometrical search algorithm is employed across all elements’ nodes and edges to figure out their configuration with respect to the discontinuity interfaces. A list consisting of all enriched elements and nodes is yielded, which in turn, is employed to initialize the XFEM physics interfaces over the enriched zone ${\Omega _{\rm{h}}}$. In the case of crack growth, the pre-processing step is repeated in each propagation step to account for the alterations in the discontinuity interfaces. This is performed by utilizing two MATLAB functions; “phi.m”, which evaluates the Heaviside function at any point of interest (e.g., Gauss points and nodes) and modifies the strain field, pressure and temperature gradients in corresponding physics interfaces, i.e., SMenr, DLenr and HTenr, respectively. The second function entitled “interpol.m”, identifies the enriched zone by means of an interpolation function. This function attributes one, to the enriched nodal points, and zero, to the remainder of nodal points. During the course of the analysis, the output variable $\psi$ is interpolated using the MATLAB internal function \textit{scatterInterpolant}, which is scripted within the external function“interpol.m”. In order to provide access to the element connectivities, nodal coordinates, interpolation values of nodal points and crack-tip coordinates at any stage of the analysis, all are stored in distinct MAT files. The above procedure is designated so as to overcome the inherent restriction of no access to the elements and nodes information encountered during the course of analysis in COMSOL. In this way, by recalling the MAT files at any stage of the analysis, the required preprocessing data is geometrically inferred from the background mesh. It is noteworthy that COMSOL allows the user to associate a single output with each MATLAB function call. As a result, the number of user-defined MATLAB functions must match the variables required during the solution. In addition, all input and output vectors associated with each of the MATLAB functions must necessarily have identical array sizes.

\textbf{Remark 1.} In certain circumstances where crack interfaces are aligned with the element edges (e.g., stationary or mode I propagating cracks), the level set updating process can be performed by relying on built-in features of COMSOL in lieu of user-defined MATLAB functions. This can be achieved by taking advantage of a series of explicit expressions that specify the bisected elements as well as the Heaviside values at Gauss integration points. This alleviates the interactions with LiveLink for MATLAB throughout the analysis and therefore, enhances the computational efficiency considerably.

\subsection{Module set-up}
\label{S:3-3 module setup}

\subsubsection{XFEM enrichment}
\label{S:3-3-1 xfem enrichment}

According to the weak formulation presented in section \ref{S:2-2 weak forms XFEM dis}, two distinct modules are set up as per physics interfaces (i.e., "Solid Mechanics", "Darcy’s Law" and "Heat Transfer in Porous Media") in order to take account of the standard and enriched terms of the corresponding field variables (i.e., $\bf{u}$, $p$ and $T$, respectively). For the continuous parts of the field variables in standard physics interfaces, no special modification is required for the definition of the strain field ${{\boldsymbol{\varepsilon }}^{{\rm{std}}}}({\bf{x}})$, pressure $\nabla {p^{{\rm{std}}}}({\bf{x}})$ and temperature gradients $\nabla {T^{{\rm{std}}}}({\bf{x}})$, thus
\begin{equation}
\label{eq:gradstdimplement}
\begin{array}{l}
{{\boldsymbol{\varepsilon }}^{{\rm{std}}}}({\bf{x}}) = {\nabla ^{\rm{s}}}{{\bf{u}}^{{\rm{c}}}}({\bf{x}})\\
\nabla {p^{{\rm{std}}}}({\bf{x}}) = \nabla {p^{{\rm{c}}}}({\bf{x}})\\
\nabla {T^{{\rm{std}}}}({\bf{x}}) = \nabla {T^{{\rm{c}}}}({\bf{x}})
\end{array}
\end{equation}
On the other hand, the enriched physics interfaces require modifications to take into account the discontinuities in the solution domain. This task is undertaken by adopting the enrichment function provided by the MATLAB function “phi.m” (i.e., $H_{\Gamma _{d}}(\varphi (\mathbf{x}))$, in here) as
\begin{equation}
\label{eq:gradenrimplement}
\begin{array}{l}
{{\boldsymbol{\varepsilon }}^{{\rm{enr}}}}({\bf{x}}) = {\nabla ^{\rm{s}}}{{\bf{u}}^{{\rm{d}}}}({\bf{x}}) \times {H_{{\Gamma _{\rm{d}}}}}(\varphi ({\bf{x}}))\\
\nabla {p^{{\rm{enr}}}}({\bf{x}}) = \nabla {p^{{\rm{d}}}}({\bf{x}}) \times {H_{{\Gamma _{\rm{d}}}}}(\varphi ({\bf{x}}))\\
\nabla {T^{{\rm{enr}}}}({\bf{x}}) = \nabla {T^{{\rm{d}}}}({\bf{x}}) \times {H_{{\Gamma _{\rm{d}}}}}(\varphi ({\bf{x}}))
\end{array}
\end{equation}
where ${{\boldsymbol{\varepsilon }}^{{\rm{enr}}}}({\bf{x}})$, $\nabla {p^{{\rm{enr}}}}({\bf{x}})$ and $\nabla {T^{{\rm{enr}}}}({\bf{x}})$ are respectively the enriched contributions of the strain field, pressure and temperature gradients. To apply such modifications, the user needs to enable “Equation View” option in COMSOL, which facilitates the access to the definition of all variables related to each physics interface.

In order to ensure the consistency of the solution across all the physics involved, further modifications need to be introduced in the definition of stress, velocity and heat flux fields. For this purpose, the overall (i.e., contributions from both standard and enriched fields) effective stress ${\boldsymbol{\bar \sigma }}$, fluid velocity $\dot{\bar{\mathbf{w}}}_{\rm{f}}$, and conductive $\bar{\mathbf{q}}_\text{cd}$ and convective $\bar{\mathbf{q}}_\text{cv}$ heat transfer vectors, which are respectively associated with Cauchy stress, Darcy’s velocity field and Conductive and Convective heat transfer vectors, are modified as
\begin{equation}
\label{eq:totalcomsol}
\begin{array}{l}
{\boldsymbol{\bar \sigma }} = {\bf{D}}:({{\boldsymbol{\varepsilon }}^{{\rm{std}}}} + {{\boldsymbol{\varepsilon }}^{{\rm{enr}}}})\\
\dot{\bar{\mathbf{w}}}_{\rm{f}} = \frac{{{k_{\rm{f}}}}}{{{\mu _{\rm{f}}}}}( - \nabla {p^{{\rm{std}}}} - \nabla {p^{{\rm{enr}}}} + {\rho _{\rm{f}}}({\bf{b}}-\ddot{\bf{ u}}))\\
\bar{\mathbf{q}}_\text{cd} = {\lambda _{{\rm{eff}}}}\nabla T = {\lambda _{{\rm{eff}}}}(\nabla {T^{{\rm{std}}}} + \nabla {T^{{\rm{enr}}}})\\
\bar{\mathbf{q}}_\text{cv} = {\rho _{\rm{f}}}{C_{\rm{f}}}\dot{\bar{\mathbf{w}}}_{\rm{f}}
\end{array}
\end{equation}
It is noteworthy that the above modifications must be identically applied to both standard and enriched sets of physics interfaces.

The enriched physics interfaces need to be merely introduced and solved over the enrichment zone ${\Omega _{\rm{h}}}$ and its associated DOFs. However, in COMSOL the physics interfaces are essentially defined across the whole domain. In order to overcome such restriction, the field variable $\psi$ (see section \ref{S:3-2 preprocessing}) is applied to prescribe the primary variables of the enriched physics interfaces such that,
\begin{equation}
{\aleph ^{{\rm{d}}}} = \left\{ {\begin{array}{*{20}{c}}
{{\aleph ^{{\rm{d}}}}}&{{\rm{where}}\,\psi  = 1}\\
0&{{\rm{otherwise}}}
\end{array}} \right.
\end{equation}
Using the above definition, the solution domain associated with the enriched physics interfaces, which is ubiquitously defined over the entire domain, can be restricted to the enriched zone. This task is performed by using the “Pointwise Constraints” from the “Domain Constraints” by selecting the same discretization settings as per their corresponding physics (e.g., shape function type, element order, etc). This, in turn, results in the elimination of the redundant DOFs that are located outside the desired enrichment zone. This strategy significantly improves the computational efficiency of the model, while problematic singularities in the stiffness matrix emanating from the incorrect definition of the enriched DOFs over the un-enriched zone are avoided.

\subsubsection{Thermo-hydro-mechanical coupling process}
\label{S:3-3-2 THM coupling}

\begin{figure}[!t]
\centering\includegraphics[width=0.6\linewidth]{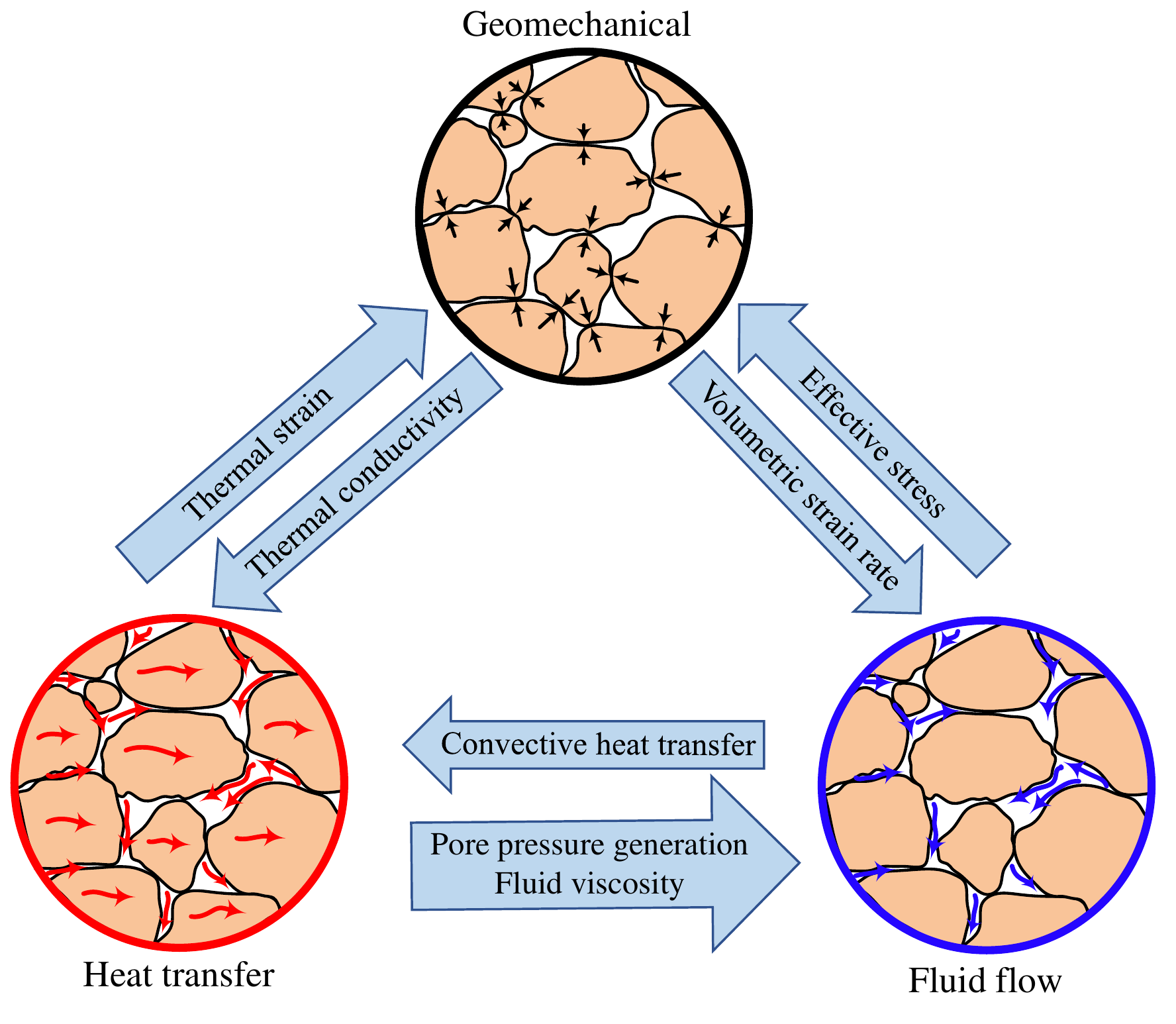}
\caption{ Thermo-hydro-mechanical coupling flowchart.}
\label{fig:THMcoupling}
\end{figure}

COMSOL Multiphysics features excellent capabilities in coupling multiple physical phenomena. This task can either be performed by selecting predefined multiphysics interfaces or by administering the coupling process manually. In the latter, the user controls the multiphysics coupling between the physics interfaces by manually specifying how each physics affects another \cite{Multiphysics2019introduction}. This technique is particularly advantageous whenever there is no preconfigured multiphysics interface designated for the physics involved, which is the case for the XFEM implementation of the thermo-hydro-mechanical coupling theme proposed here. Each of the physics involved (i.e., geomechanical deformation, fluid flow and heat transfer processes) essentially interacts with another, which in turn generates a coupled multiphysics theme as demonstrated in Fig. \ref{fig:THMcoupling}.

\textit{- Thermo-mechanical coupling:}

\noindent Thermally induced strain is accommodated within the Solid Mechanics physics interface by the use of the “Thermal Expansion” subnode \cite{multiphysics2019structural}. It is noteworthy that this term must only be incorporated into the SMstd physics interface, to ensure the thermal strains are applied consistently over the entire strain field.

\textit{- Hydro-mechanical coupling:}

\noindent The poroelastic coupling terms are defined in accordance with Biot's poroelasticity theory \cite{multiphysics2019subsurface}. In this sense, the effective stress principle is implemented via the “External Stress” subnodes in both SMstd and SMenr, where the “stress input” option is set to the “pore pressure”. Moreover, the variations of the volumetric strain, that directly contribute to the changes in pore space (i.e., the first term in Eq. \ref{eq: flow-continuity}), are included by means of “Mass Source” subnodes in the DLstd and DLenr physics interfaces.

\textit{- Hydro-thermal coupling:}

\noindent The hydro-thermal coupling theme is automatically captured by activating the “Nonisothermal Flow” within HTstd and HTenr physics interfaces, with no need to introduce any additional flow and/or heat coupling terms manually \cite{multiphysics2019heat}. 

\subsubsection{Thermo-mechanical contact}
\label{S:3-3-3 contact}

\begin{figure}
\centering
\begin{subfigure}{.5\textwidth}
  \centering
  \includegraphics[width=.7\linewidth]{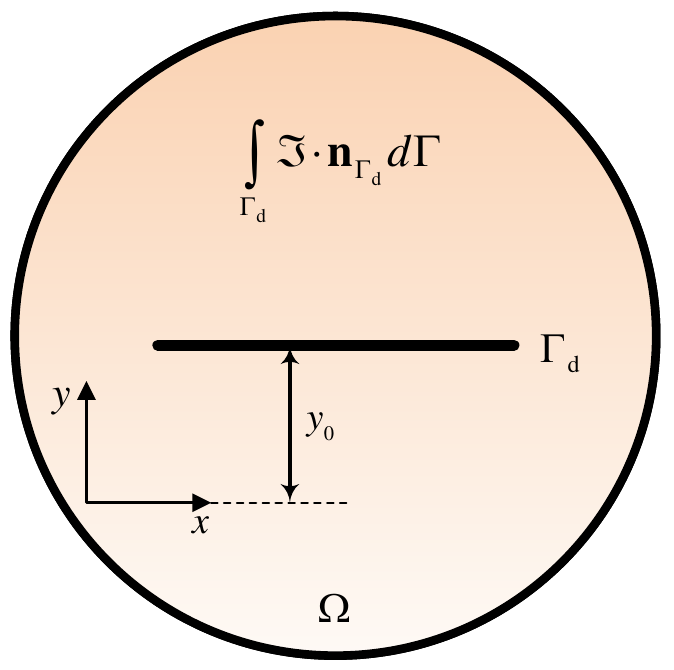}
  \caption{}
  \label{fig:Dirac1}
\end{subfigure}%
\begin{subfigure}{.5\textwidth}
  \centering
  \includegraphics[width=.7\linewidth]{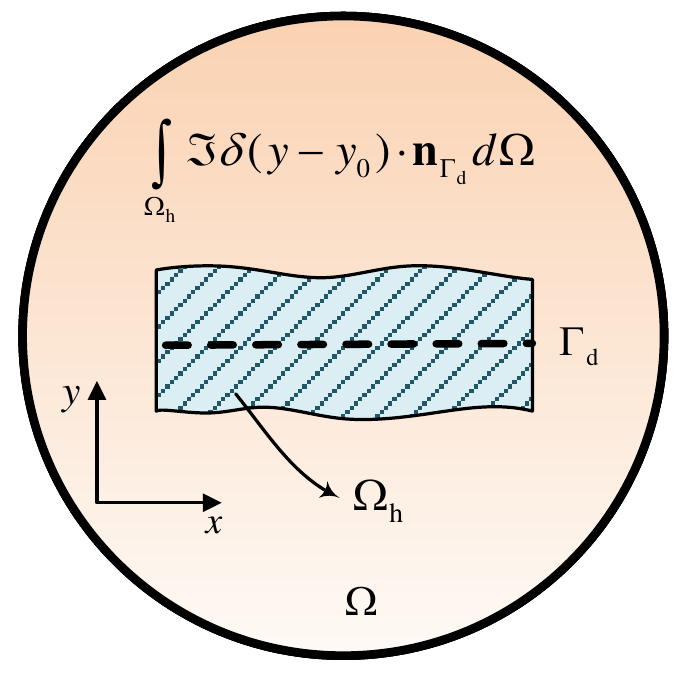}
  \caption{}
  \label{fig:Dirac2}
\end{subfigure}%
\caption{a) Numerical line integration of an arbitrary function $\Im$ along the discontinuity; b) Its equivalent domain integral over the enriched zone by using Dirac delta function $\delta$.}
\label{fig:Dirac}
\end{figure}

COMSOL Multiphysics is endowed with mechanical and thermal contact modeling algorithms. However, these built-in capabilities can be elaborated only on pre-determined contact surfaces, which need to be explicitly introduced in the problem configuration. To facilitate the imposition of XFEM contact constraints, the $\psi$ field is employed such that the solution domain is delimited to the enriched zone ${\Omega _{\rm{h}}}$, as explained in section \ref{S:3-2 preprocessing}. Nevertheless, in COMSOL Multiphysics, it is not possible to perform a numerical integration over a non-predefined geometrical contact surface. This restriction is circumvented by adding a domain “Weak Contribution” node in SMenr and HTenr physics interfaces, whereas a penalty contact algorithm is elaborated to account for contact traction and heat flux in accordance to Kuhn-Tucker inequalities (i.e., Eq. \ref{eq: kuhn-tucker}). Moreover, the line integral of the contact contributions along the contact surface is converted to an equivalent domain integration over the enriched zone $\Omega_{\rm{h}}$ through the introduction of a special Dirac delta function $\delta$ over the discontinuity interface. This process is illustrated schematically in Fig. \ref{fig:Dirac}, which pertains to the case of a horizontal discontinuity interface, supposedly, at a vertical distance $y_0$ from the origin as
\begin{equation}
\label{eq:Dirac}
\int\limits_{{\Gamma _{\rm{d}}}} {\Im \cdot {{\bf{n}}_{{{\Gamma}}_{\rm{d}}}} d\Gamma }  = \int\limits_{{\Omega _{\rm{h}}}} {\Im \delta (y - {y_0}) \cdot {{\bf{n}}_{{{\Gamma}}_{\rm{d}}}}d\Omega }
\end{equation}
where $\Im$ is the normal contact force/heat flux. In this way, the domain integration over ${{\Omega _{\rm{h}}}}$ (i.e., 2D/3D) is equivalent to the line integral along the crack dimension ${{\Gamma _{\rm{d}}}}$ (i.e., 1D/2D). 

\subsection{Numerical integration}
\label{S:3-4 integration}

The piece-wise continuous polynomials that are employed in the conventional FEM are accurately integrated by means of lower-order Gauss quadrature rules. In contrast, due to the presence of various discontinuities and/or singularities in the primary field variables and, possibly, their corresponding derivatives, XFEM requires more precise techniques to enable the accurate evaluation of integrals over the enriched region. In this regard, high-order Gauss integration schemes, rectangular sub-griding and element partitioning (e.g., triangular/rectangular) are most frequently employed in fracture analysis \cite{khoei2014extended}. Nevertheless, only the first approach is viable in COMSOL and, therefore, it is applied for the numerical simulations. 

\textbf{Remark 2.} In order to ensure that an ill-posed and/or singular stiffness matrix is avoided in the XFEM analysis, a sufficient number of Gauss integration points must exist at either side of the discontinuity interfaces. To this end, an additional criterion is introduced so as to examine the presence of a minimum size of the support domain associated with each node, for a certain integration order. Accordingly, a node is enriched only if its support domain i.e., $A^{+}/(A^{+}+A^{-})$ or $A^{-}/(A^{+}+A^{-})$ exceeds a predefined tolerance $\delta_{\rm{s}}$, where $A^{+}$ and $A^{-}$ are the bisected areas of each element formed by the discontinuity interface. This criterion, evidently, avoids the mathematical redundancy of the solution \cite{mohammadi2008extended}.

\subsection{Solution procedures and analysis strategies}
\label{S:3-5 analysis}

The process of solving a discretized boundary value problem in COMSOL is a hierarchy that is executed from the “Study” node \cite{multiphysics2019refmanual}. From several available study options in COMSOL, “Stationary” and “Time Dependent” subnodes are among the frequently used ones for the simulation of porous media, that pertain to the nature of the numerical analysis (i.e., quasi-static/steady-state and dynamic/transient, respectively). Whenever studying quasi-static crack propagation analysis, the prescribed load is applied incrementally. Meanwhile, the crack interfaces are deemed to propagate once a certain crack growth criterion is met. This task is performed in COMSOL by using the “Auxiliary Sweep” section of the “Stationary” subnode, whereas the applied incremental load is taken as the sweep parameter. In contrast, the “Time Dependent” study option is inherently history-dependent and thus, the sweep strategy is not applicable. For further detail on the XFEM implementation of crack growth in COMSOL, refer to \cite{jafari2021XFEM}.

COMSOL provides two options for solving the nonlinear system of equations i.e., fully coupled and segregated schemes. Depending on the nonlinearity of the equations and the degree of coupling, the methods can be employed interchangeably which, in turn, provides great flexibility for multiphysics problems in terms of accuracy and efficiency. In this study,  the fully coupled implicit solver is applied in conjunction with \say{Generalized alpha} time integration method, which is a finite difference based scheme, to execute the transient time-dependent thermo-hydro-mechanical analysis. The step-by-step implementation procedure adopted for the XFEM thermo-hydro-mechanical modeling in COMSOL is concisely presented in Algorithm \ref{alg:localmins}, which involves: global/local variables and geometry definitions, selection of standard and enriched physics interfaces along with the associated modifications, domain discretization and mesh generation, and numerical solution technique. 
\begin{algorithm}[hbt!]
\caption{Step-by-step implementation of XFEM thermo-hydro-mechanical model in COMSOL.} 
\label{alg:localmins}
\begin{algorithmic}
\STATE 1. Global Definitions
\STATE \hspace{10mm} Define all constants (material, load, etc)
\STATE \hspace{10mm} Define MATLAB functions (\say{phi.m}, \say{interpol.m}, etc)
\STATE 2. Create Geometry (2D/3D)
\STATE 3. Local Variables definition
\STATE \hspace{10mm} Define interaction integral equations
\STATE \hspace{10mm} For propagating cracks: Assign Global Variable Probe; call MATLAB functions for crack update,
\STATE \hspace{10mm} crack tip and angle (see reference \cite{jafari2021XFEM})
\STATE 4. Select physical model (physics interfaces)
\STATE \hspace{10mm} Standard Solid Mechanics (SMstd)
\STATE \hspace{17mm} Select material model
\STATE \hspace{17mm} Modify stress definitions (Eq. \ref{eq:totalcomsol})
\STATE \hspace{17mm} Add thermo-mechanical coupling using Thermal Expansion subnode
\STATE \hspace{17mm} Add hydro-mechanical coupling (effective stress) using External Stress subnode

\STATE \hspace{10mm} Enriched Solid Mechanics (SMenr)
\STATE \hspace{17mm} Select material model
\STATE \hspace{17mm} Modify strain definitions (Eq. \ref{eq:gradenrimplement})
\STATE \hspace{17mm} Modify stress definitions (Eq. \ref{eq:totalcomsol})
\STATE \hspace{17mm} Add hydro-mechanical (effective stress) coupling
\STATE \hspace{17mm} Apply field variable $\psi$ as a constraint using the domain Pointwise Constraint option

\STATE \hspace{10mm} Standard Darcy's Law (DLstd)
\STATE \hspace{17mm} Select fluid and matrix properties
\STATE \hspace{17mm} Modify Darcy's velocity definitions (Eq. \ref{eq:totalcomsol})
\STATE \hspace{17mm} Add hydro-mechanical coupling (rate of volumetric strain) using Mass Source subnode

\STATE \hspace{10mm} Enriched Darcy's Law (DLenr)
\STATE \hspace{17mm} Select fluid and matrix properties
\STATE \hspace{17mm} Modify Darcy's velocity definitions (Eq. \ref{eq:totalcomsol})
\STATE \hspace{17mm} Modify pressure gradient definitions (Eq. \ref{eq:gradenrimplement})
\STATE \hspace{17mm} Add hydro-mechanical coupling (rate of volumetric strain) using Mass Source subnode
\STATE \hspace{17mm} Apply field variable $\psi$ as a constraint using the domain Pointwise Constraint option

\STATE \hspace{10mm} Standard Heat Transfer in Porous Media (HTstd)
\STATE \hspace{17mm} Select porous medium and fluid properties
\STATE \hspace{17mm} Activate Nonisothermal Flow
\STATE \hspace{17mm} Modify Conductive heat flux definitions (Eq. \ref{eq:totalcomsol})
\STATE \hspace{17mm} Modify Convective heat flux definitions (Eq. \ref{eq:totalcomsol})

\STATE \hspace{10mm} Enriched Heat Transfer in Porous Media (HTenr)
\STATE \hspace{17mm} Select porous medium and fluid properties
\STATE \hspace{17mm} Activate Nonisothermal Flow
\STATE \hspace{17mm} Modify Conductive heat flux definitions (Eq. \ref{eq:totalcomsol})
\STATE \hspace{17mm} Modify Convective heat flux definitions (Eq. \ref{eq:totalcomsol})
\STATE \hspace{17mm} Modify heat gradient definitions (Eq. \ref{eq:gradenrimplement})
\STATE \hspace{17mm} Apply field variable $\psi$ as a constraint using domain Pointwise Constraint option

\STATE 5. Assign initial and boundary conditions
\STATE 6. Domain discretization and mesh generation
\STATE 7. Specify Study type
\STATE \hspace{10mm} Select Time Dependent (transient) or Stationary (steady-state) analysis
\STATE 8. Post-processing and visualization

\end{algorithmic}

\end{algorithm}
 \section{Numerical simulations}
\label{S:4 (results)}

Several numerical examples are presented in order to illustrate the robustness and performance of the proposed XFEM implementation procedure. The first example deals with the validation of the solution strategy in handling temperature-induced fracture growth in a bi-material composite beam. A thermo-mechanical analysis is next performed on a cracked domain, for which the stress intensity factors (SIFs)and temperature distribution are evaluated and compared to the available analytical solutions in the literature. The capabilities of the proposed XFEM implementation are further elaborated in the thermo-mechanical contact analysis conducted on a doubled clamped beam consisting of a middle crack. Subsequently, an XFEM hydro-mechanical modeling of impermeable discontinuities in saturated porous media is presented. The final set of examples is devoted to illustrating the rigor of the proposed strategy in dealing with challenging thermo-hydro-mechanical simulations of 2D/3D deformable porous rocks in the presence of single/multiple faults.

Unless otherwise specified, linear elastic material behaviour and plane strain condition are stipulated for all 2D numerical simulations, where bilinear quadrilateral elements are utilized to discretize the domain. A $20 \times 20$ Gauss quadrature is applied over the enriched elements (i.e., enriched physics interfaces) to achieve satisfactory integration accuracy. In all cases, $\delta_s$ is set to 0.005 to ensure a sufficient number of Gauss integration points are available at either side of the discontinuities inside the enriched elements.

\subsection{Temperature induced fracturing of a notched bi-material beam}
\label{S:4-1 (experimental)}
\begin{figure}[!t]
\centering\includegraphics[width=0.85\linewidth]{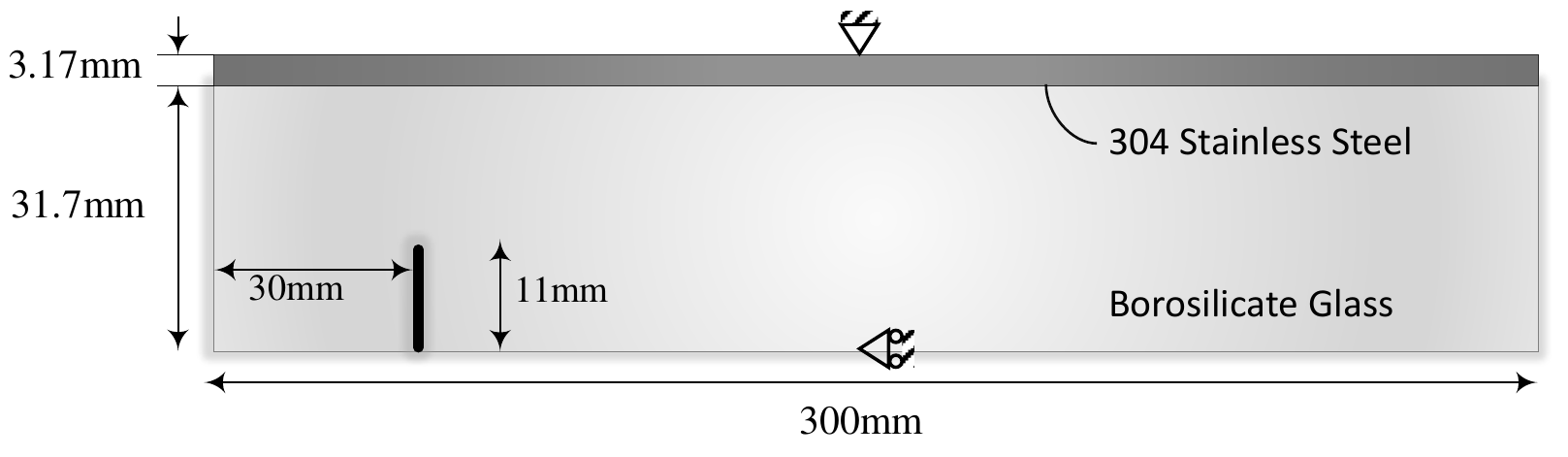}
\caption{The geometry and boundary conditions of the borosilicate/steel composite beam. }
\label{fig:ex-1 geometry}
\end{figure}

The first example studies crack growth in a borosilicate/steel bi-material beam under thermal loading, which was originally introduced and experimentally examined in \cite{grutzik2018crack}. As shown in Fig. \ref{fig:ex-1 geometry}, the composite beam consists of a strip of AISI 304 stainless steel glued to the top edge of a bar of borosilicate glass, with the respective dimensions of $3.17 \text{mm} \times 300 \text{mm}\times6\text{mm}$ and $31.7\text{mm}\times300\text{mm}\times6\text{mm}$. An initial notch is placed at a distance of 30 mm from the left edge of the specimen. The mechanical and thermal material properties of the domain are presented in Table \ref{t:ex-1}. Although geometrically simple, the specimen presents a mechanistically complex scenario regarding temperature-induced crack propagation analysis.

\begin{table}[ht]
\caption{Material properties for the bi-material beam problem.}
\centering 
\begin{tabular}{c c c c c c }
\hline \hline 
  & $E$ (GPa)  &  $\nu $ &  $\alpha_{T}$ (1/$^ \circ $C)& $G_{\rm{f}}$ (N/mm) &  $f_{\rm{t}}$ (MPa) \\ [0.5ex]
\hline                  
Borosilicate glass &	64  &  0.2  & $3.25\times10^{-6}$  & 0.4 &	80  \\
304 Stainless steel	&   193  &  0.29 & $17.3\times10^{-6}$ & -   &    -    \\ [1ex]      
\hline
\end{tabular}
\label{t:ex-1}
\end{table}

According to the experiment procedure, the specimen is cooled down from the initial room temperature (i.e.,  $20^\circ $C). The sample undergoes pure bending induced by the difference in the thermal expansion coefficients of its constituent materials. Observations show that a crack is initiated at the notch tip (i.e., within the brittle glass material), which eventually propagates towards the bi-material interface. Midway through, this trend is altered by the diversion of the crack propagation trajectory in a direction parallel to the glass/steel interface. This problem is modeled in COMOSL by using 14,432 elements under plane stress condition. The average element size is maintained at $0.5$ mm in the vicinity of the notch zone. For the sake of simplicity, the material interface is assumed to be fully bonded (no-slip constraint). The initial crack length and growth increment are set to 11 mm and $2.5$ mm, respectively. The thermal loading is applied by imposing a uniform temperature reduction over the entire domain which, in turn, results in uniform thermal strain ${{\boldsymbol{\varepsilon }}_{\rm{T}}} = \alpha \Delta T{\bf{I}}$ , where $\Delta T=-1\times i$, with  $i= 1,..,15$ being the load increment multiplier. Quasi-static crack growth analysis, using the maximum hoop stress criterion of $f_{\rm{t}}=80 \text{MPa}$, is performed through the “Auxiliary Sweep" feature in the “Extended Study" section of the “Stationary" node in COMSOL (for further detail on quasistatic crack growth in COMSOL, see \cite{jafari2021XFEM}). Note the mixed-mode SIFs are evaluated following the procedures explained in \ref{S:6 Appendix}.

\begin{figure}
\centering

\begin{subfigure}{.5\textwidth}
  \centering
  \includegraphics[width=.9\linewidth]{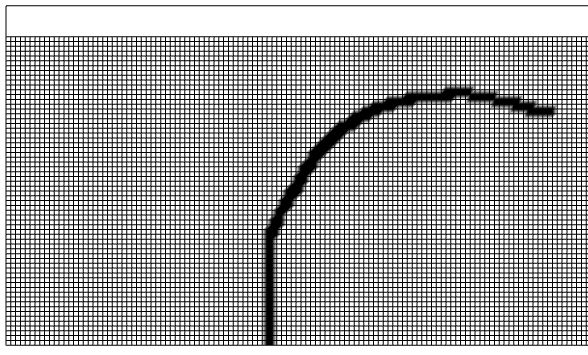}
  \caption{}
  \label{fig:ex0-pathcomsol}
\end{subfigure}%
\begin{subfigure}{.5\textwidth}
  \centering
  \includegraphics[width=.92\linewidth]{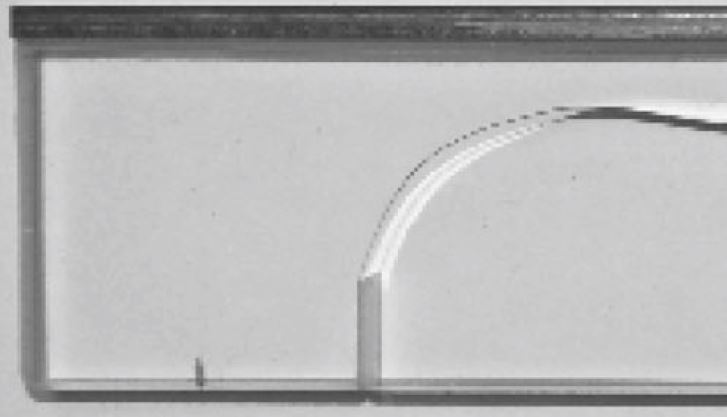}
  \caption{}
  \label{fig:ex0-pathexp}
\end{subfigure}%

\caption{Crack propagation in the bi-material beam; a) numerical crack trajectory based on $\psi$, b) experimental crack pattern \cite{grutzik2018crack}.}
\label{fig:ex0-path}
\end{figure}

Figs. \ref{fig:ex0-pathcomsol} and \ref{fig:ex0-pathexp} respectively show the simulated and experimental crack propagation trajectory for the composite specimen. The former is determined using the field variable $\psi$, that is equal to unity for fully fractured elements. As can be seen, there is a good agreement between the proposed numerical solution and the experiment. 

\subsection{Plate with an edge crack subjected to thermal loading }
\label{S:4-2 (thermomechanical)}

\begin{figure}[!t]
\centering\includegraphics[width=0.25\linewidth]{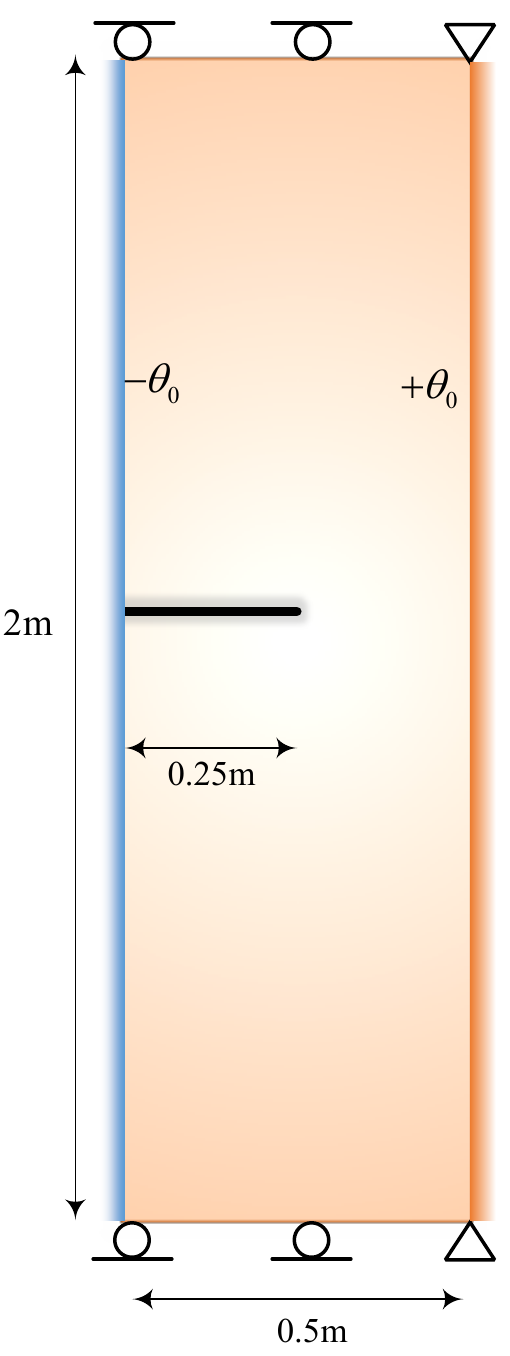}
\caption{Geometry and boundary conditions of the rectangular plate with an edge crack. }
\label{fig:ex-2 geometry}
\end{figure}

\begin{figure}\centering
\subfloat[]{\label{fig:ex2-SIF}\includegraphics[width=.5\linewidth]{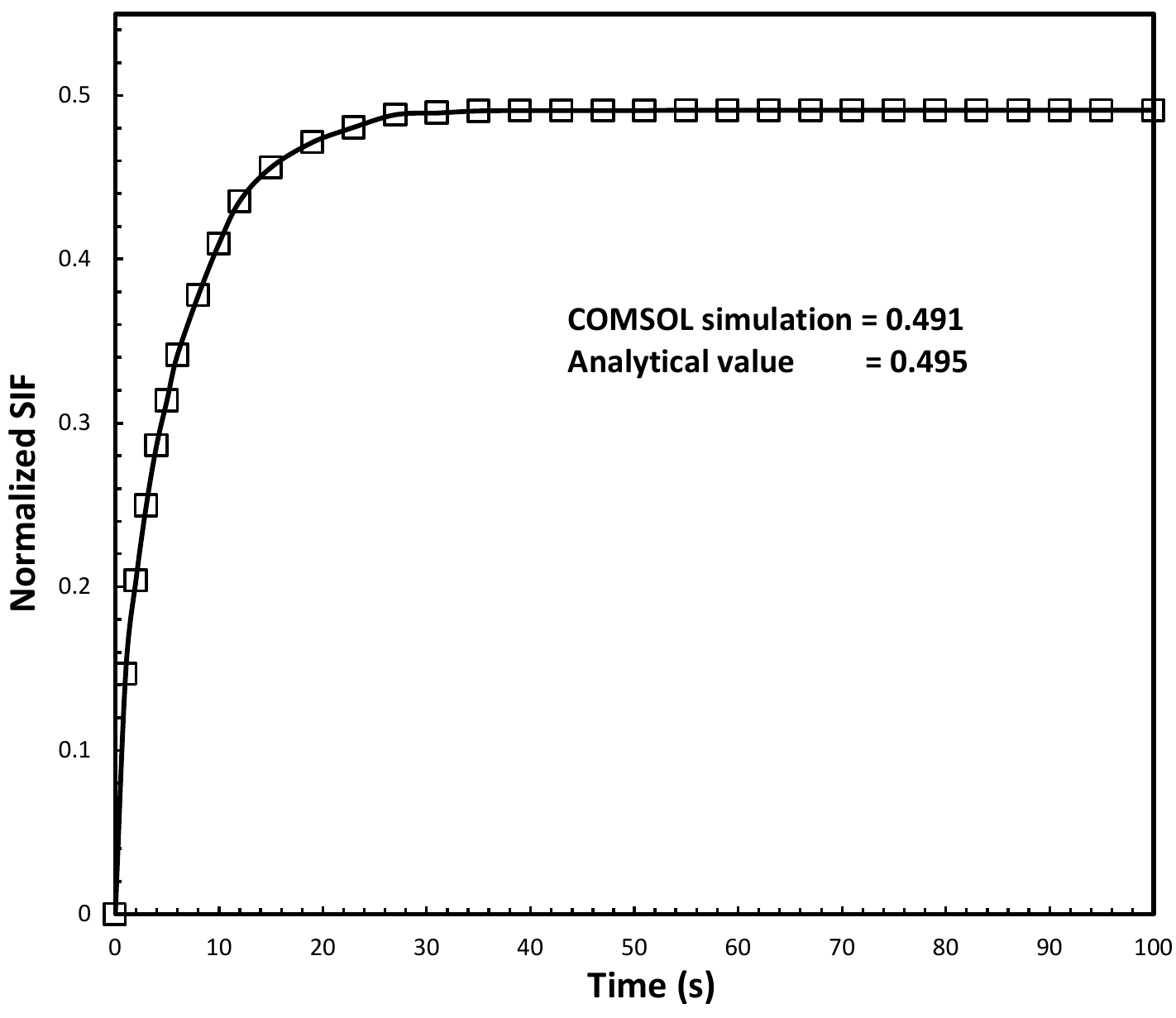}}\hfill
\subfloat[]{\label{fig:ex2-graphtemp}\includegraphics[width=.5\linewidth]{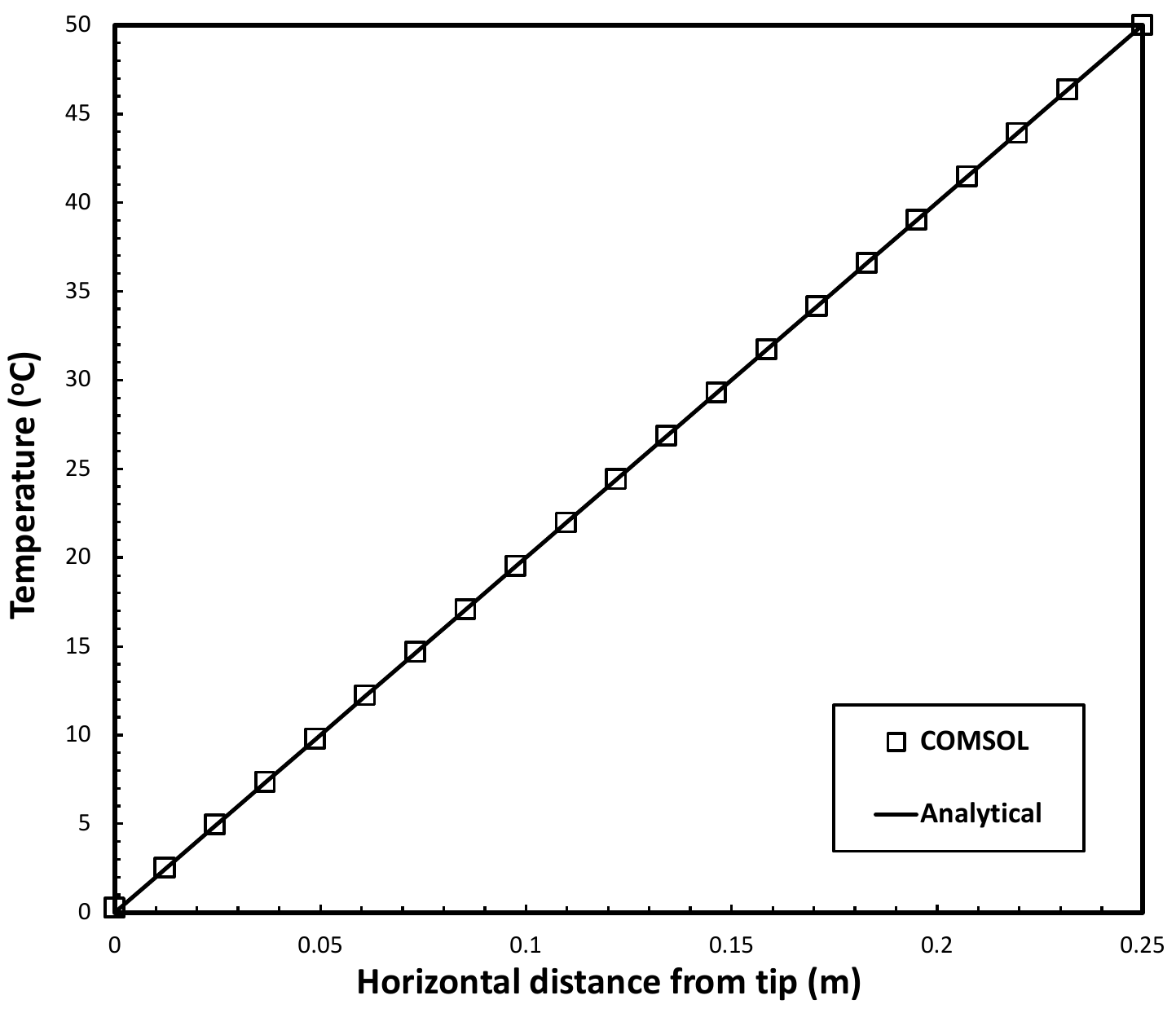}}\par 
\caption{Thermo-mechanical analysis of edge cracked plate; a) the evolution of normalized SIF and b) comparison of the COMSOL and analytical temperature distributions with the distance from crack tip. }
\label{fig:ex2-1}
\end{figure}

In the second example, a rectangular plate consisting of an adiabatic edge crack is considered under thermal loading (Fig. \ref{fig:ex-2 geometry}). This problem was first introduced by Wilson et al. \cite{wilson1979use,shih1986energy}, which presents an analytical study for an isolated crack in thermo-elastic media. This example was later exploited by Duflot \cite{duflot2008extended} to develop an XFEM formulation for the modeling of thermo-elastic fractures. Here, the reference solution is particularly employed to illustrate the accuracy of the evaluated SIF for the non-isothermal cracked settings using the proposed XFEM strategy. The problem involves a 2.0 m$\times$0.5 m plate possessing a half-width sized adiabatic edge crack. The plate is subjected to prescribed temperatures of $+50^{\rm{o}}{\rm{C}}$ and $-50^{\rm{o}}{\rm{C}}$ on the right and left edges, respectively. The thermo-mechanical properties of the plate are presented in Table \ref{t:ex2}. A structured uniform mesh of $160\times40$ elements is supposed, which is used for the coupled thermo-mechanical analysis with a total duration of $100$ s. Based on Duflot \cite{duflot2008extended}, the normalized stress intensity factor $F_{\rm{I}}$ is evaluated as
\begin{equation}
\label{eq:normalSIF}
{F_{\rm{I}}} = \frac{{{K_{\rm{I}}}}}{{\frac{E}{{1 - \nu }}\alpha {\theta _0}\sqrt {\pi a} }}
\end{equation}
where $K_{\rm{I}}$ is the mode I SIF, $a$ is the crack length, and $\theta _0$ is the absolute value of the applied temperature on the plate boundary. The time variation of the computed normalized SIF is compared with the analytic solution in Fig. \ref{fig:ex2-SIF}. It is observed that the non-isothermal solution reaches the steady-state value of 0.491 in 20 s, which indicates less than $1\%$ relative error with respect to the exact analytical value of 0.495 \cite{shih1986energy}.

\begin{table}[ht]
\caption{Material properties for the plate with an edge crack under thermal loading problem.}
\centering 
\begin{tabular}{c c  }
\hline               
\hline
Young's modulus, $E$ (GPa) &	9    \\
Poisson's ratio, $\nu$	&   0.3  \\
Solid density, $\rho_{\rm{s}}$ (kg/m$^{3}$)  & $2\times10^{3}$ \\
Thermal conductivity, $\lambda_{\rm{s}}$ (W/m $^ \circ $C)  & $1\times10^{3}$ \\
Volumetric thermal expansion coefficient, $\beta_{\rm{s}}$ (1/$^ \circ $C) & $3\times10^{-7}$\\ [0.5ex]      
\hline
\end{tabular}
\label{t:ex2}
\end{table}

\begin{figure}
\centering
\begin{subfigure}{.33\textwidth}
  \centering
  \includegraphics[width=.8\linewidth]{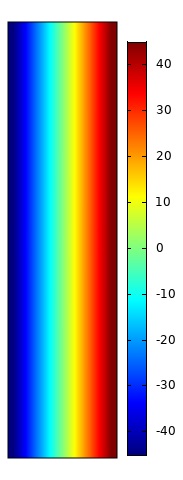}
  \caption{}
  \label{fig:ex2-temp}
\end{subfigure}%
\begin{subfigure}{.33\textwidth}
  \centering
  \includegraphics[width=0.8\linewidth]{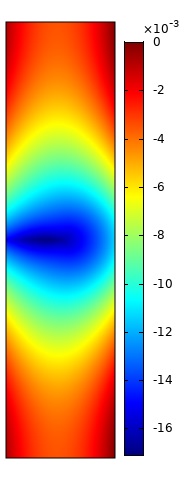}
  \caption{}
  \label{fig:ex2-ux}
\end{subfigure}
\begin{subfigure}{.33\textwidth}
  \centering
  \includegraphics[width=0.8\linewidth]{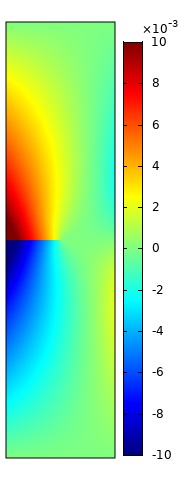}
  \caption{}
  \label{fig:ex2-uy}
\end{subfigure}
\caption{The final distribution contours of a) temperature ($^ \circ {\rm{C}}$), b) horizontal displacement (mm), and c) vertical displacement (mm). }
\label{fig:ex2-2}
\end{figure}

Fig. \ref{fig:ex2-graphtemp} denotes the profile of temperature variation versus the distance from the crack tip. The simulation results are in an excellent agreement with the exact solution expressed as $\theta (x) = (2x - w){\theta _0}$ \cite{duflot2008extended}. Here, $\theta (x)$ is the temperature at coordinate $x$ with respect to the crack tip, and $w$ is the plate’s width. The contours of the temperature, vertical and horizontal displacements are also depicted in Fig. \ref{fig:ex2-2}. Evidently, an intensified contraction of the crack edges can be detected, which is attributed to the loss of strength along the crack faces.

\subsection{Thermo-mechanical analysis of a cracked double-clamped beam}
\label{S:4-3 (thermalcontact)}

\begin{figure}[!t]
\centering\includegraphics[width=0.85\linewidth]{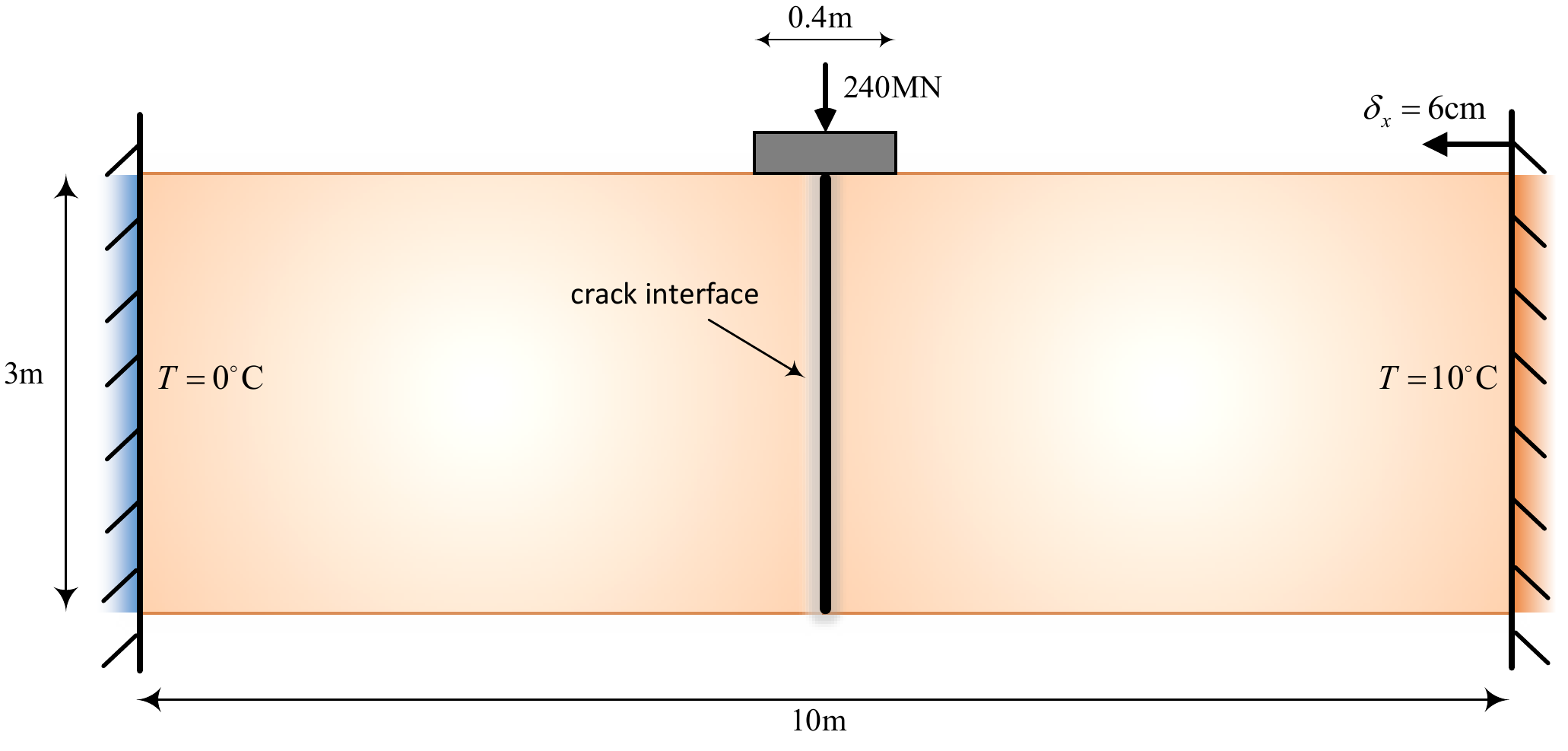}
\caption{Geometry and boundary conditions of the double-clamped beam under thermo-mechanical loading. }
\label{fig:ex-3 geometry}
\end{figure}

The following example explores the thermo-mechanical contact analysis in a double clamped beam under simultaneous thermal and mechanical loading. This problem was initially introduced by Khoei \cite{khoei2014extended} for purely mechanical contact analysis. As shown in Fig. \ref{fig:ex-3 geometry}, the domain consists of a 3 m $\times$ 10 m beam, containing a full depth vertical crack at the middle. The beam is restrained along its left end, and is subjected to a prescribed uniform horizontal displacement of $6$ cm at the opposite end. The left edge is also subjected to a prescribed temperature of 0 $^{\rm{o}}{\rm{C}}$,  while the right edge is retained at a constant temperature of 10 $^{\rm{o}}{\rm{C}}$. Furthermore, a vertical loading of $240$ MN is applied along the top edge of the beam, which is distributed uniformly over a distance of $0.4$ m about its center. The thermo-mechanical material properties used in this example are summarized in Table \ref{t:ex3}. The domain is discretized using a structured mesh consisting of $155\times47$ quadrilateral elements. As a result of the presence of the full depth vertical crack, the beam is divided into two distinct segments. Considering the symmetry of the problem, no relative vertical displacement (i.e., slip) may occur between the crack faces. Therefore, only the normal contact constraint is applied by means of a penalty approach, with no frictional resistance. A normal penalty parameter of ${10^9}$ ${\rm{MN/}}{{\rm{m}}^{\rm{3}}}$ is selected for the mechanical phase. To impose a fully continuous temperature field across the closure zone (i.e., active contact zone), a thermal conductivity coefficient of ${h_{{\rm{cont}}}}={10^5}$ W/m$^ \circ $C is considered within the thermal contact formulation. 

\begin{table}[ht]
\caption{Material properties for the double-clamped beam.}
\centering 
\begin{tabular}{c c  }
\hline               
\hline
Young's modulus, $E$ (GPa) &	10    \\
Poisson's ratio, $\nu$	&   0.3  \\
Density, $\rho_{\rm{s}}$ (kg/m$^{3}$)  & $10^{3}$ \\
Thermal conductivity, $\lambda_{\rm{s}}$ (W/m $^ \circ $C)  & $150$ \\
Specific heat capacity, C$_{\rm{s}}$ (J/kg $^ \circ $C) & 100 \\ 
Volumetric thermal expansion coefficient, $\beta_{\rm{s}}$ (1/$^ \circ $C) & $23.86\times10^{-6}$\\ [0.5ex]      
\hline
\end{tabular}
\label{t:ex3}
\end{table}

\begin{figure}
\centering
\begin{subfigure}{.45\textwidth}
  \centering
  \includegraphics[width=.85\linewidth]{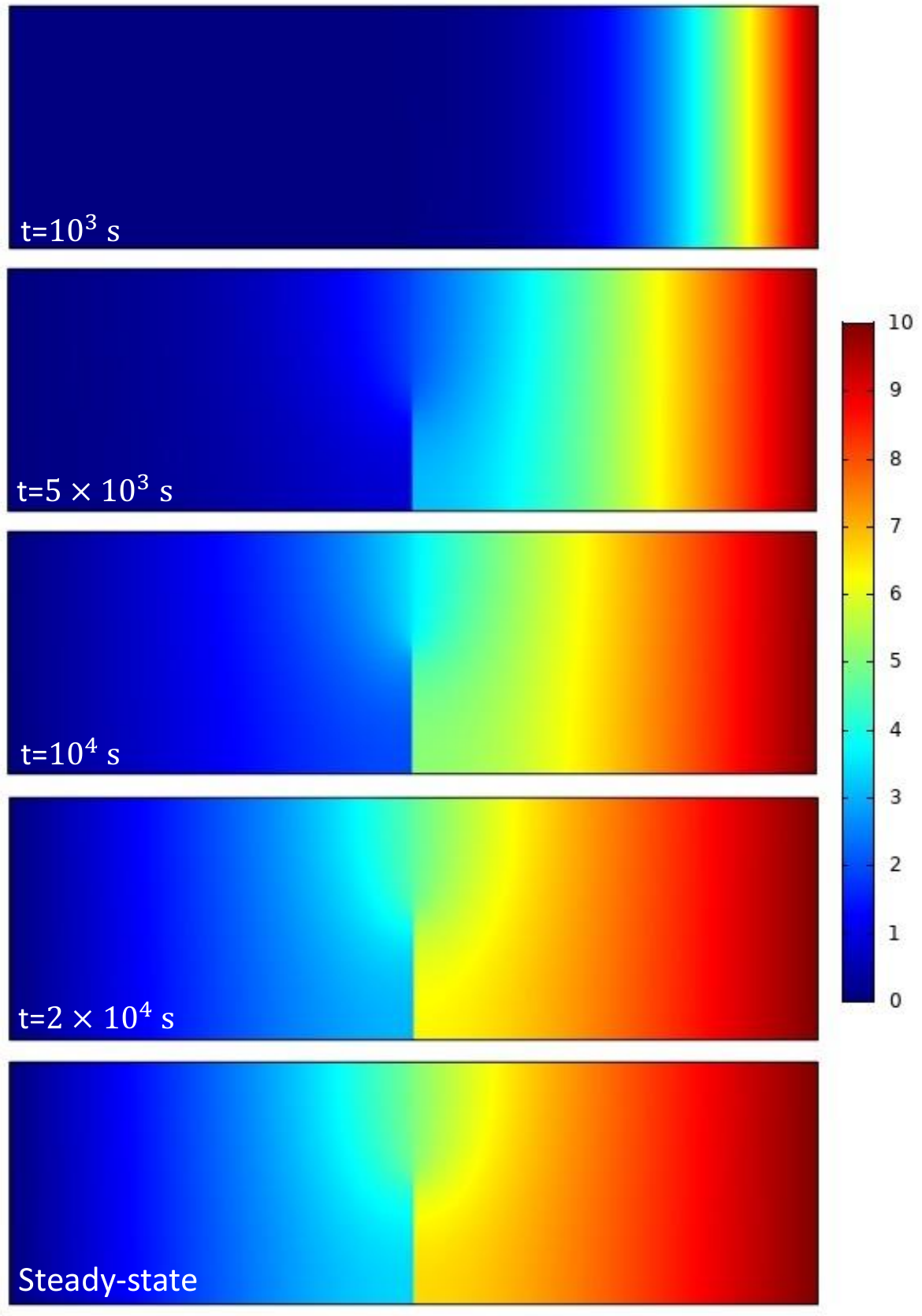}
  \caption{}
  \label{fig:ex3-temp}
\end{subfigure}%
\begin{subfigure}{.45\textwidth}
  \centering
  \includegraphics[width=0.85\linewidth]{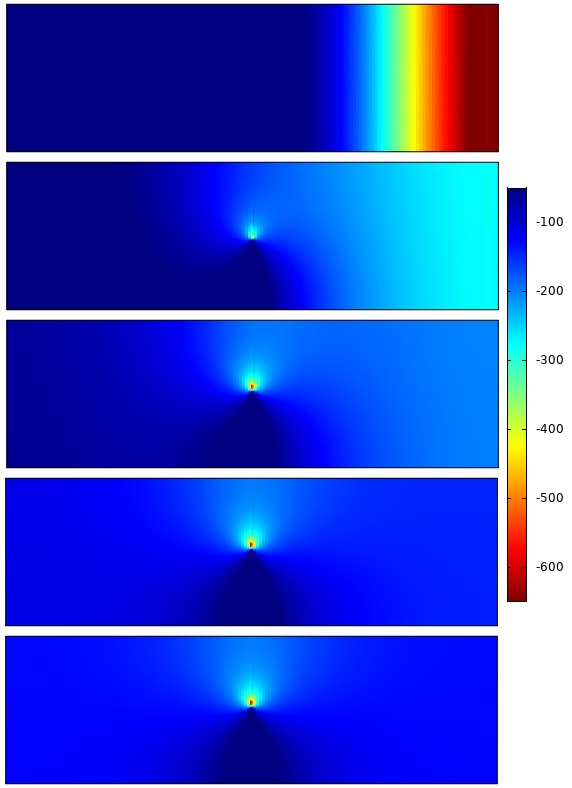}
  \caption{}
  \label{fig:ex3-flux}
\end{subfigure}
\begin{subfigure}{.45\textwidth}
  \centering
  \includegraphics[width=0.85\linewidth]{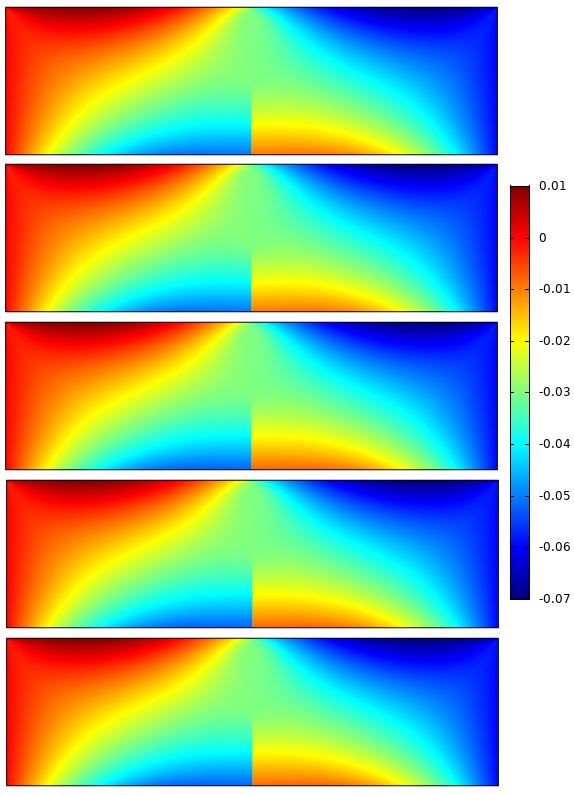}
  \caption{}
  \label{fig:ex3-ux}
\end{subfigure}
\begin{subfigure}{.45\textwidth}
  \centering
  \includegraphics[width=0.85\linewidth]{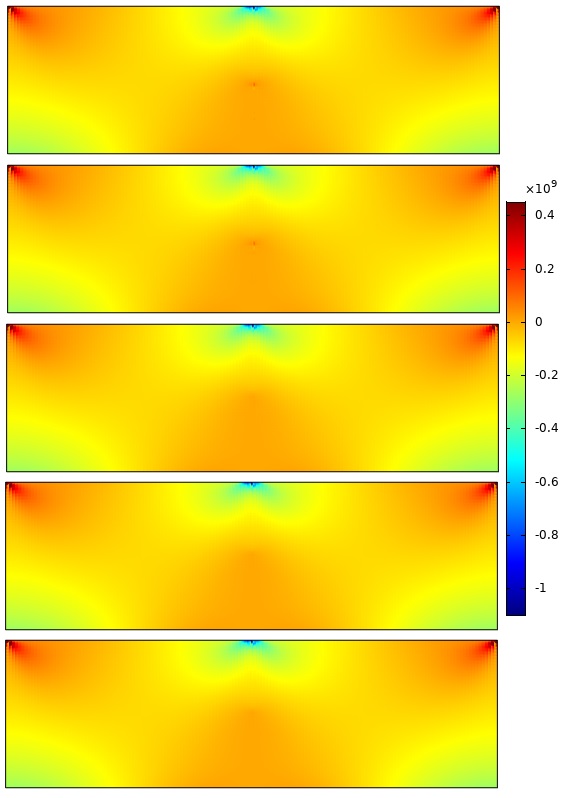}
  \caption{}
  \label{fig:ex3-sxx}
\end{subfigure}
\caption{The contours of a) temperature ($^ \circ {\rm{C}}$), b) horizontal heat flux (${\rm{N/s}}{\rm{.m}}$), c) horizontal displacement (m), and d) normal stress in $x$ direction ${\sigma _x}$ (Pa) at $10^{3}$s, $5\times10^{3}$s, $10^{4}$s, $2\times10^{4}$s and the steady-state condition (from top to bottom).  }
\label{fig:ex-3 contours}
\end{figure}

\begin{figure}[!t]
\centering\includegraphics[width=0.49\linewidth]{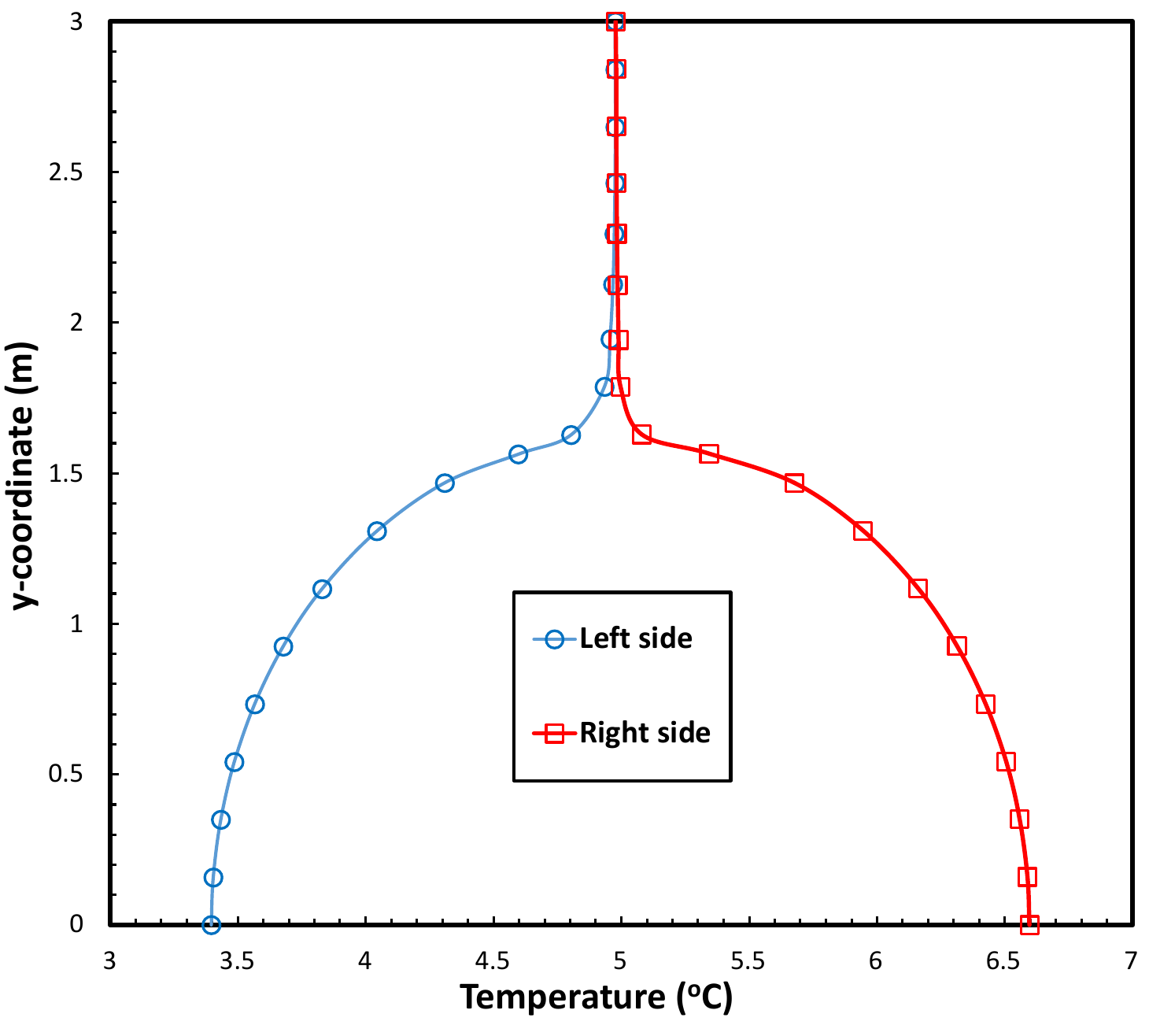}
\caption{Temperature profiles along the both sides of the discontinuity at the end of simulation. }
\label{fig:ex-3 tempgraph}
\end{figure}

In this example, the active contact zone is not known a \textit{priori}; alternatively, it is determined following an iterative solution procedure which mitigates the non-positive values of the normal gap (i.e., $\left[\kern-0.15em\left[ {{u_N}} 
 \right]\kern-0.15em\right]$=0). As explained in section \ref{S:3-3-3 contact}, this task is executed by using the enriched portion of the displacement field, which is explicitly derived from SMenr. Both mechanical and thermal contact contributions are applied via exploiting the domain \say{Weak Contribution} subnodes that are integrated using a $20 \times 20$ Gauss integration quadrature. The use of higher order Gauss quadrature ensures that the integration of the contact force/heat flux along crack faces is accurately performed by using their corresponding equivalent domain integrals. Fig. \ref{fig:ex-3 contours} depicts the distribution of temperature, horizontal heat flux $q_x$, the horizontal displacement $u_x$, and normal stress in $x$ direction $\sigma_{xx}$, at various stages of the time-dependent analysis. Fig. \ref{fig:ex3-flux} illustrates a continuous and increasing heat flux across the active contact zone with a significant flux concentration in the vicinity of the crack-tip. The results also demonstrate the formation of a continuous temperature field over the contact zone, particularly as the simulation reaches the steady-state condition. This can be observed likewise in Fig. \ref{fig:ex-3 tempgraph}, where the temperature profiles are plotted along both crack faces. As predicted, in opening mode, the Heaviside enrichment function has induced a jump in the temperature field.

\subsection{Hydro-mechanical modeling of impermeable discontinuities in an embankment dam}
\label{S:4-4 (hydromechanical)}

\begin{figure}[!t]
\centering\includegraphics[width=0.85\linewidth]{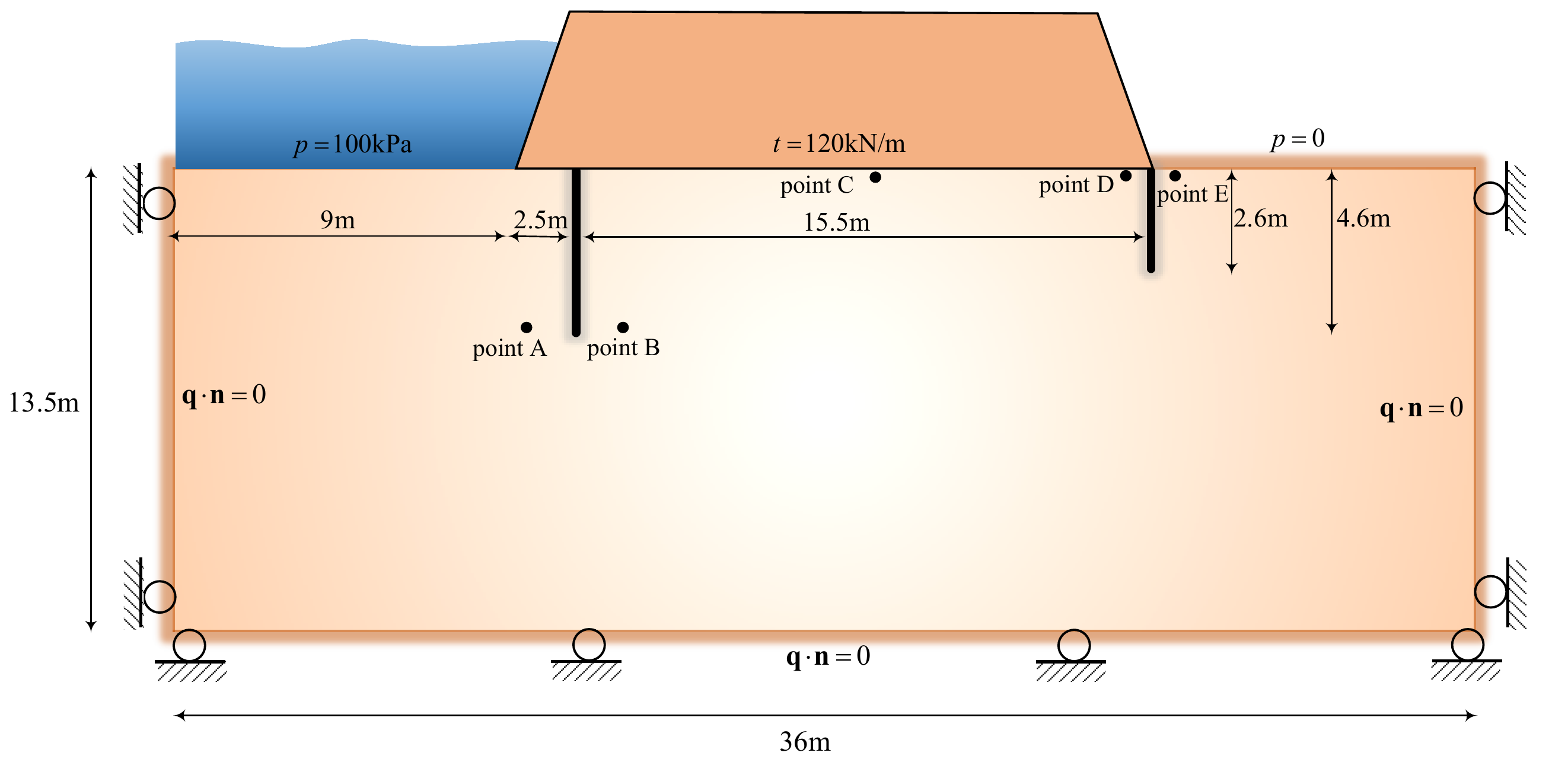}
\caption{Embankment dam problem with impermeable discontinuities: geometry and boundary conditions. }
\label{fig:ex-4 geometry}
\end{figure}

This example investigates the applicability of the proposed framework in dealing with the hydro-mechanical simulation of impermeable discontinuities in porous media. Sheet piles, frequently used in dam engineering practice, are structural barriers inserted beneath embankment dams to increase the pore water pressure dissipation. This, in turn, reduces the uplift pressure applied to the dam base as well as the occurrence of sand boiling phenomenon (a.k.a. quicksand) downstream of the dam \cite{das2019advanced}. In this example, using the XFEM, an impermeable discontinuity is introduced to emulate the presence of sheet piles on seepage flow in a porous medium. As illustrated in Fig. \ref{fig:ex-4 geometry}, a 13.5 m $\times$ 36 m silty sand formation is considered which is subjected to the vertical uniform surcharge of $12$ ton/m from the embankment dam as well as a $100$ kPa pressure from the upstream water reservoir. Two vertical sheet piles with lengths of $4.6$ m and $2.6$ m are inserted beneath the dam at a lateral spacing of $15.5$ m apart from each other. The downstream water level is supposed to be at the ground level (i.e., $p=0$). To investigate the effect of the mesh size on the solution accuracy, the domain is discretized by three different mesh configurations using the average element sizes of $0.5$ m, $0.25$ m and $0.17$ m. The time-dependent analysis is performed for the period of $110$ s. The mechanical and hydraulic properties used for this study are provided in Table \ref{t:ex4}.

\begin{table}[ht]
\caption{Material properties for the embankment dam problem containing impermeable discontinuities.}
\centering 
\begin{tabular}{c c  }
\hline               
\hline
Young's modulus, $E$ (GPa) &	9    \\
Poisson's ratio, $\nu$	&   0.4  \\
Solid density, $\rho_{\rm{s}}$ (kg/m$^{3}$)  & $2\times10^{3}$ \\
Fluid density, $\rho_{\rm{f}}$ (kg/m$^{3}$)  & $10^{3}$ \\
Porosity, n  & 0.3 \\
Bulk modulus of solid, $K_{\rm{s}}$ (MPa)  & $1\times10^{14}$ \\
Bulk modulus of fluid, $K_{\rm{f}}$ (MPa)  & $2\times10^{3}$ \\
Fluid viscosity, $\mu_{\rm{f}}$ (Pa.s)  & $2\times10^{-3}$ \\
Permeability, $k_{\rm{f}}$ (m$^{2}$) & $1\times10^{-12}$\\ [0.5ex]      
\hline
\end{tabular}
\label{t:ex4}
\end{table}

\begin{figure}[!t]
\centering\includegraphics[width=0.7\linewidth]{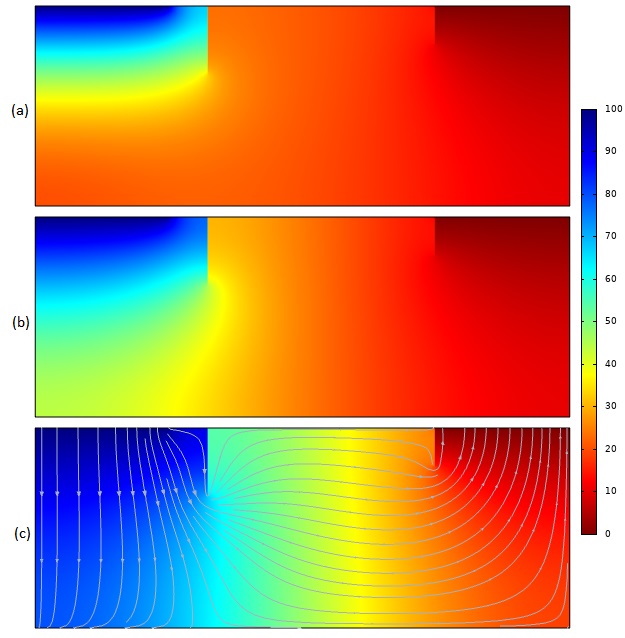}
\caption{The pressure distribution contours at 5 s, 15 s and 110 s (values are in kPa). }
\label{fig:ex-4 pressure}
\end{figure}

\begin{figure}
\centering
\begin{subfigure}{.5\textwidth}
  \centering
  \includegraphics[width=1.0\linewidth]{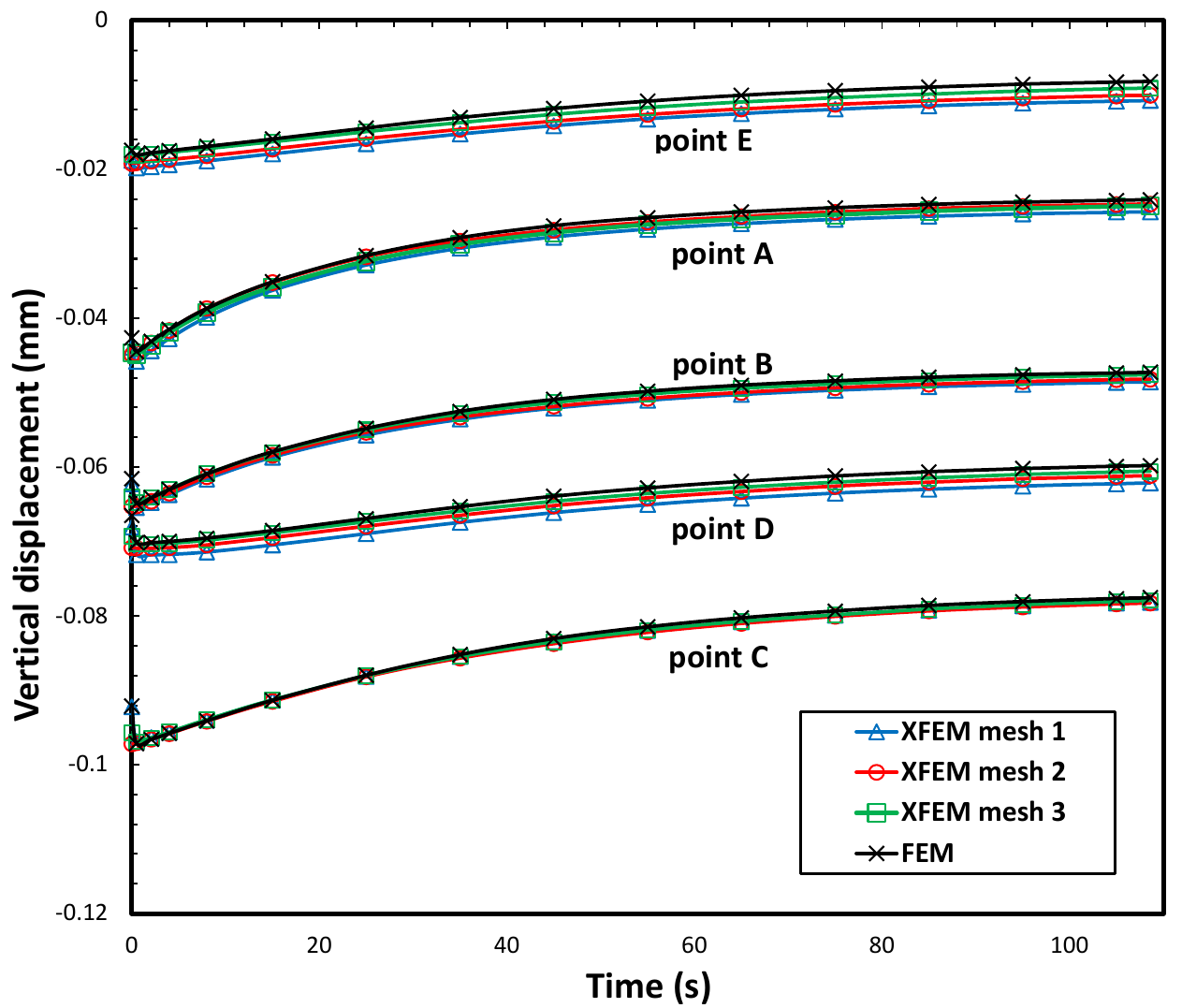}
  \caption{}
  \label{fig:ex4-disp}
  
\end{subfigure}%
\begin{subfigure}{.5\textwidth}
  \centering
  \includegraphics[width=.98\linewidth]{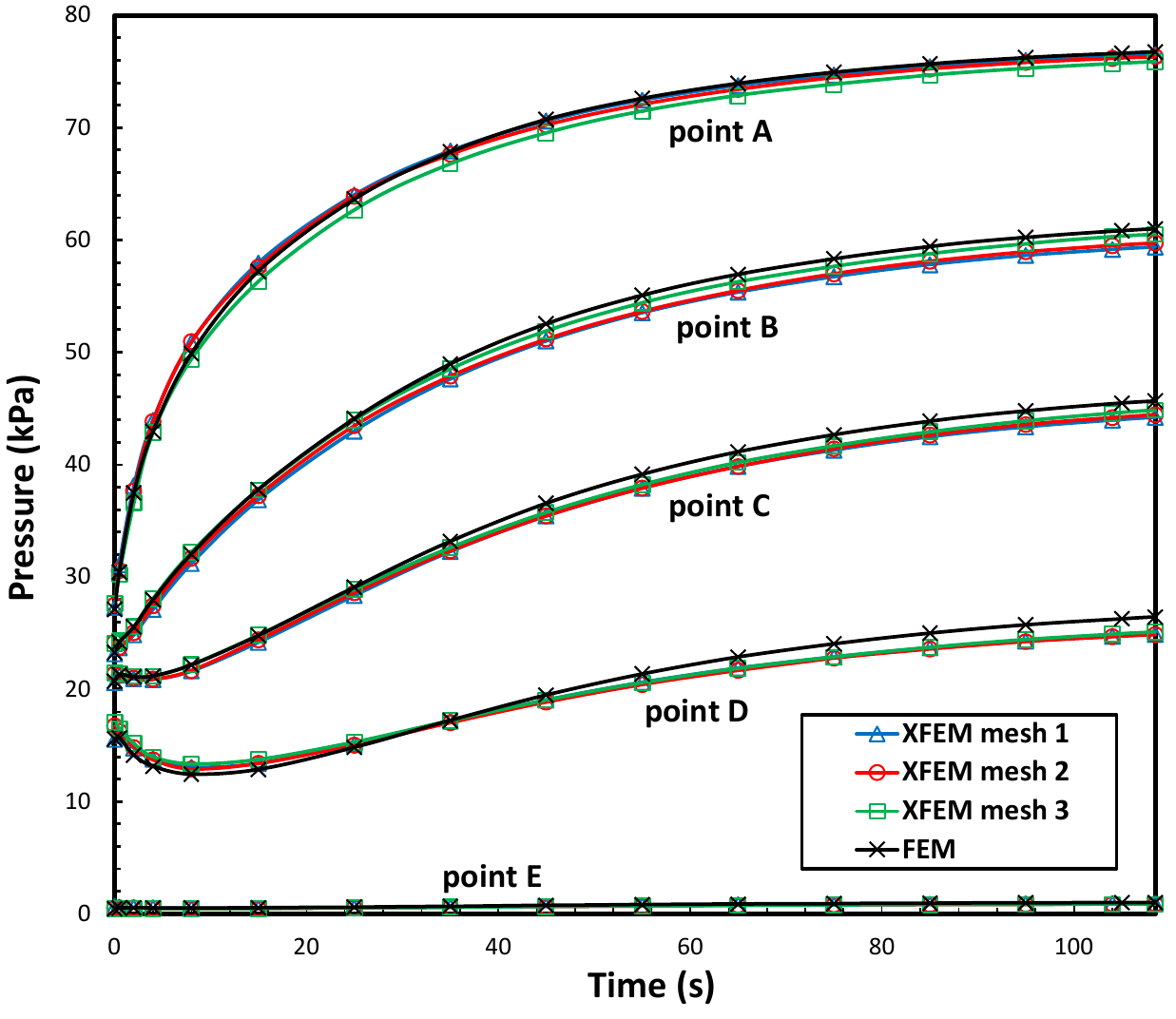}
  \caption{}
  \label{fig:ex4-pressure}
\end{subfigure}%

\caption{The evolution of a) vertical displacement, and b) pressure at various critical points in the porous domain; comparison of proposed XFEM with the classic FEM strategy}
\label{fig:ex4-plots}
\end{figure}

Fig. \ref{fig:ex-4 pressure} illustrates the evolution of pressure at three instants of the analysis i.e., $5$ s, $15$ s and $110$ s. In the latter case (Fig. \ref{fig:ex-4 pressure}c), the flow streamlines are also depicted to highlight the effects of the sheet piles on the deviation of the flow path, particularly along the dam-domain interface zone. To further elaborate on this issue, the time variation of vertical displacement and pore water pressure are plotted in Fig. \ref{fig:ex4-plots} for five critical points A, B, C, D and E, which are positioned on either side of the sheet pile as well as the midpoint beneath the dam body (indicated in Fig. \ref{fig:ex-3 geometry}). The results obtained by the classic finite element analysis are also shown in Fig. \ref{fig:ex4-plots} for the sake of comparison. Notably, the FE analysis is performed by stipulating that the sheet piles are narrow 2D devoid with impermeable boundaries. According to Fig. \ref{fig:ex4-plots}, an excellent agreement is observed between the proposed XFEM strategy and the reference FEM solution for both displacement and pressure fields. Meanwhile, the convergence of the simulation results corresponding to the three different mesh configurations is deemed satisfactory.

\subsection{Thermo-hydro-mechanical simulation of faults in 2D/3D porous media}
\label{S:4-5 (THM) 2D}

Thermo-hydro-mechanical modeling of fractures and faults manifest in a wide range of geomechanics engineering applications, including geothermal energy extraction, oil and gas recovery and deep geological disposal of nuclear wastes \cite{najari2014thermo}. The following numerical simulations are presented to demonstrate the remarkable applicability of the proposed framework in dealing with the 2D/3D thermo-hydro-mechanical analysis of impermeable faults in porous media. The thermo-hydro-mechanical material properties elaborated in this study are presented in Table \ref{t:ex5}.

\begin{table}[ht]
\caption{Material properties for the thermo-hydro-mechanical analysis of faults in the porous medium.}
\centering 
\begin{tabular}{c c  }
\hline               
\hline
Young's modulus, $E$ (GPa) &	1.6    \\
Poisson's ratio, $\nu$	&   0.33  \\
Solid density, $\rho_{\rm{s}}$ (kg/m$^{3}$)  & $2\times10^{3}$ \\
Fluid density, $\rho_{\rm{f}}$ (kg/m$^{3}$)  & $10^{3}$ \\
Porosity, n  & 0.3 \\
Bulk modulus of solid, $K_{\rm{s}}$ (MPa)  & $1\times10^{14}$ \\
Bulk modulus of fluid, $K_{\rm{f}}$ (MPa)  & $2\times10^{3}$ \\
Fluid viscosity, $\mu_{\rm{f}}$ (Pa.s)  & $2\times10^{3}$ \\
Permeability, $k_{\rm{f}}$ (m$^{2}$) & $1\times10^{-12}$\\
Thermal conductivity of solid, $\lambda_{\rm{s}}$ (W/m $^ \circ $C)  & 2.88 \\
Thermal conductivity of fluid, $\lambda_{\rm{f}}$ (W/m $^ \circ $C)  & 0.6 \\
Solid specific heat capacity, C$_{\rm{s}}$ (J/kg $^ \circ $C) & $1.17\times10^{3}$ \\
Fluid specific heat capacity, C$_{\rm{f}}$ (J/kg $^ \circ $C) & $4.2\times10^{3}$ \\
Volumetric thermal expansion coefficient, $\beta_{\rm{s}}$ (1/$^ \circ $C) & $6.6\times10^{-6}$ \\ [0.5ex]      
\hline
\end{tabular}
\label{t:ex5}
\end{table}

\subsubsection{Inclined fault in 2D porous media}
\label{S:4-5-1 (THM)}

\begin{figure}[!t]
\centering\includegraphics[width=0.5\linewidth]{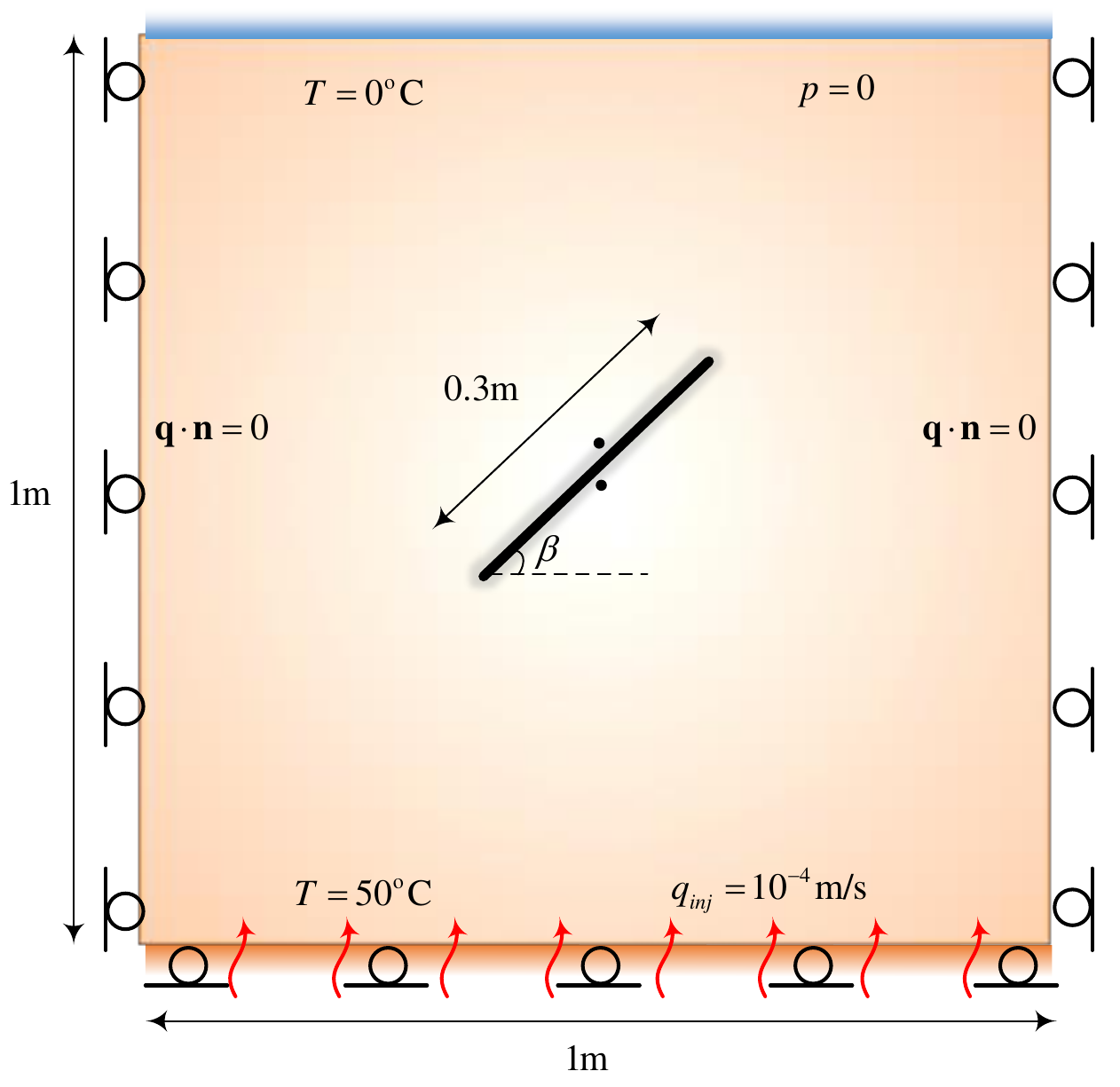}
\caption{Thermo-hydro-mechanical modeling of an inclined fault: geometry and boundary conditions.}
\label{fig:ex-5-1 geometry}
\end{figure}

The first problem investigates the effects of the presence of an inclined impermeable fault, with the length of 0.3 m and inclination angle of $\beta$ with respect to the horizon, on the thermo-hydro-mechanical response of a 1 m $\times$ 1 m square porous block (Fig. \ref{fig:ex-5-1 geometry}). The bottom edge of the block is subjected to a prescribed temperature of 50 $^{\rm{o}}{\rm{C}}$ and normal fluid flux of $10^{-4}$ m/s. The top edge is assumed to be in a drained condition, where both pressure and temperature vanish (i.e., $p=0$; $T=0$). The domain is discretized by using a structured mesh of $91\times91$ elements. Various fault angles i.e., $\beta$=0, $30^{\rm{o}},45^{\rm{o}},60^{\rm{o}}$ and $90^{\rm{o}}$, are used to study the significance of strong discontinuity onto both pressure and temperature fields, over a total period of $10^{4}$ s.

\begin{figure}\centering
\subfloat[]{\label{fig:ex5-1-disp}\includegraphics[width=.49\linewidth]{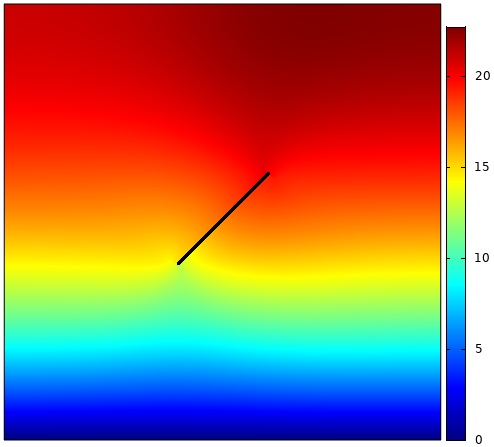}}\hfill
\subfloat[]{\label{fig:ex5-1-pressure}\includegraphics[width=.49\linewidth]{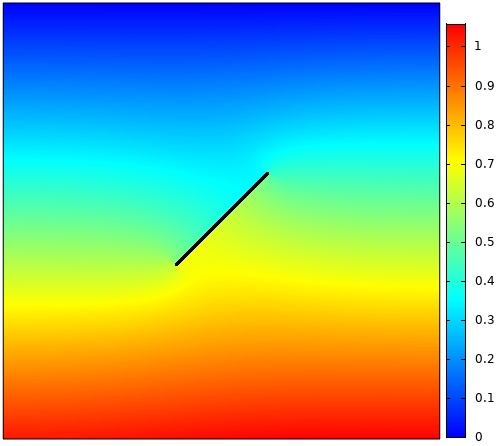}}\par 
\subfloat[]{\label{fig:ex5-1-temp}\includegraphics[width=.49\linewidth]{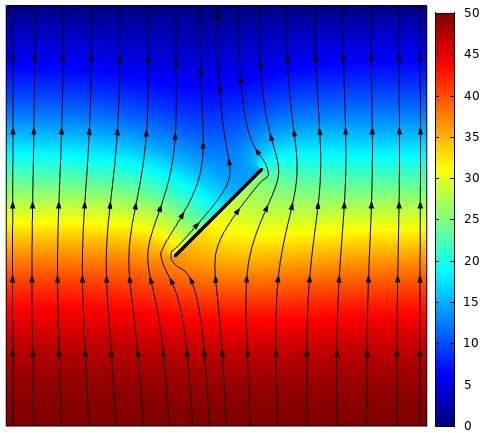}}
\caption{The distribution contours of a) vertical displacement (mm), b) pressure (MPa) and c) temperature ($^ \circ {\rm{C}}$) at the end of the analysis.  }
\label{fig:ex5-1contours}
\end{figure}

\begin{figure}\centering
\subfloat[]{\label{fig:ex5-1-pbot}\includegraphics[width=.49\linewidth]{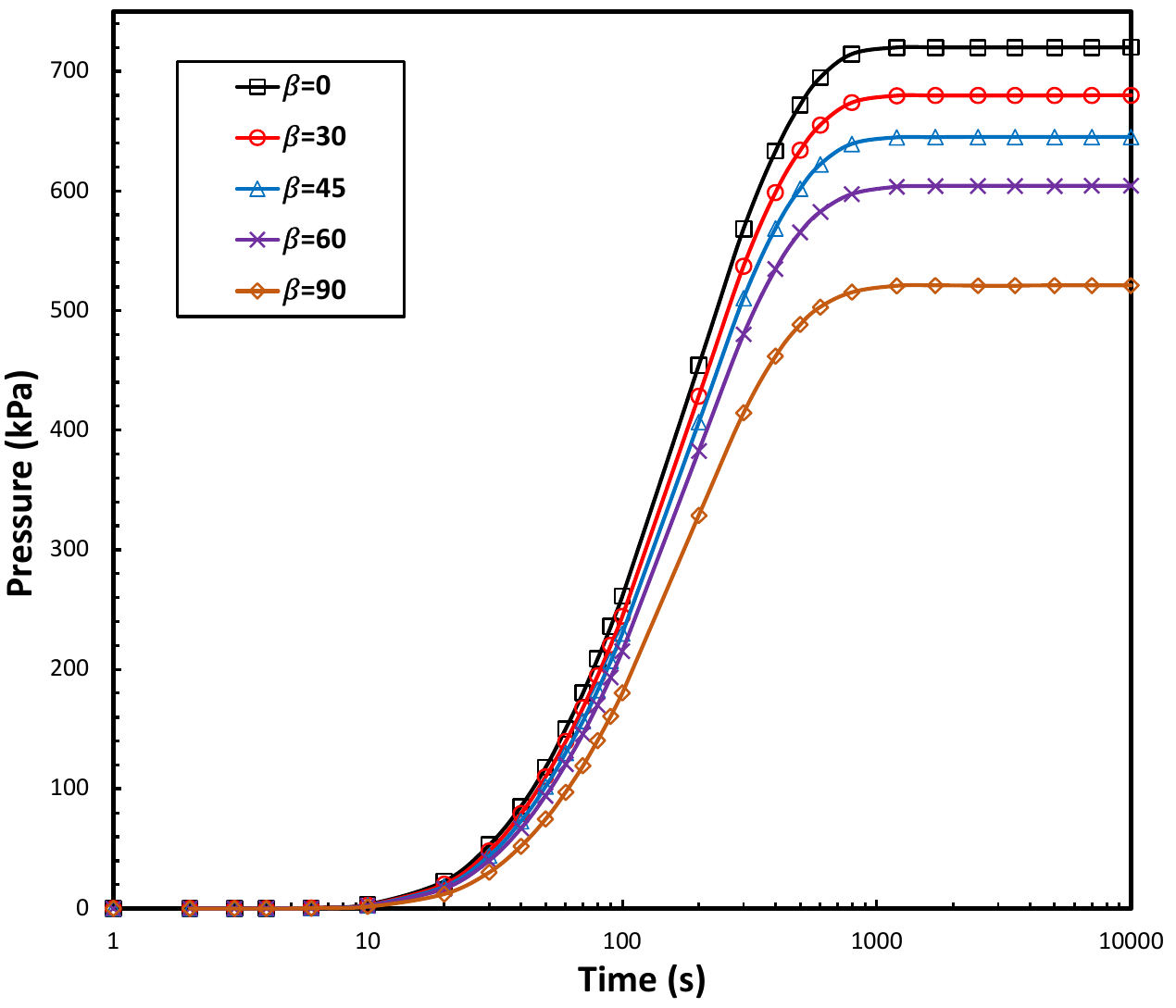}}\hfill
\subfloat[]{\label{fig:ex5-1-ptop}\includegraphics[width=.49\linewidth]{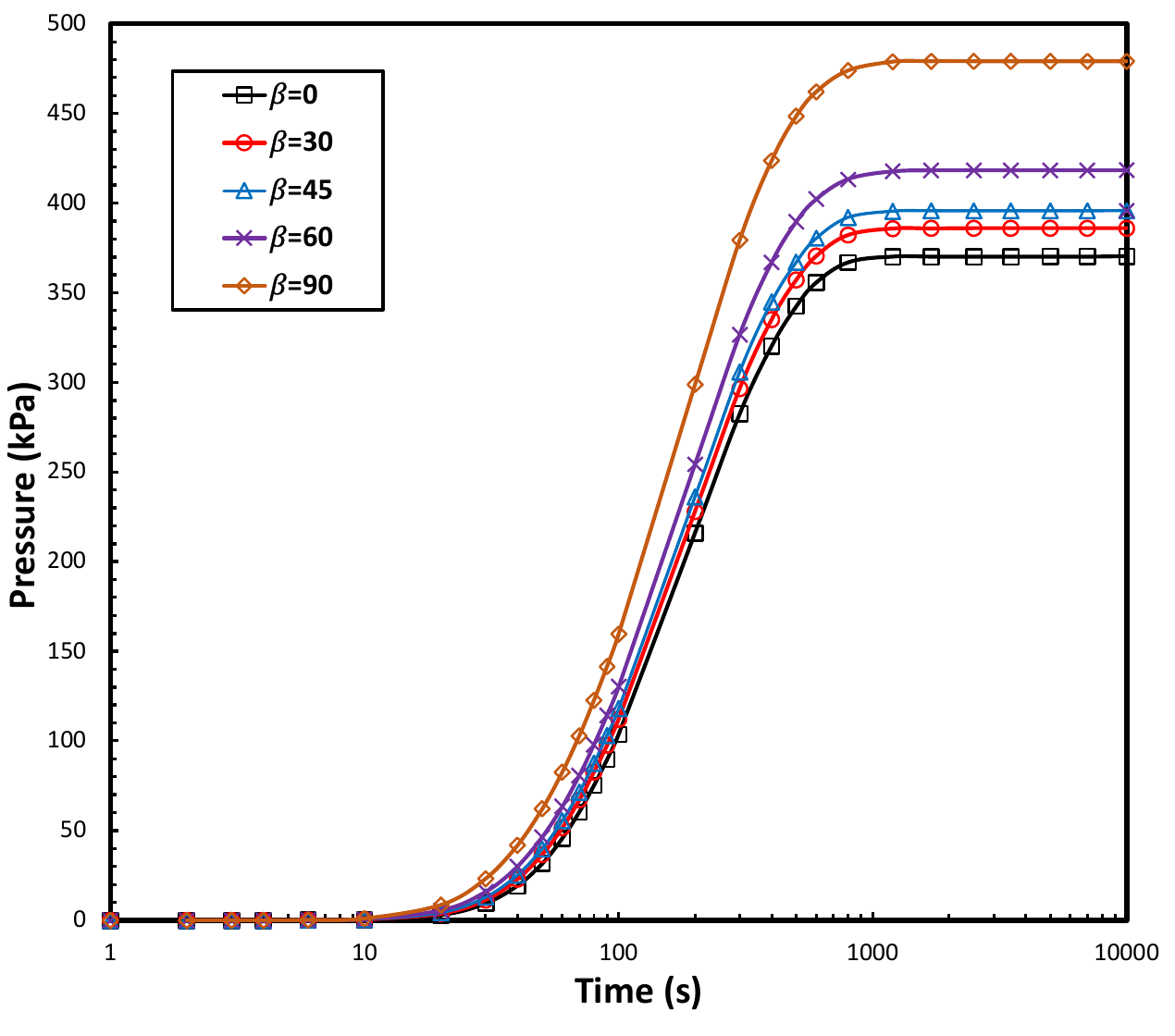}}\par 
\subfloat[]{\label{fig:ex5-1-pdiff}\includegraphics[width=.49\linewidth]{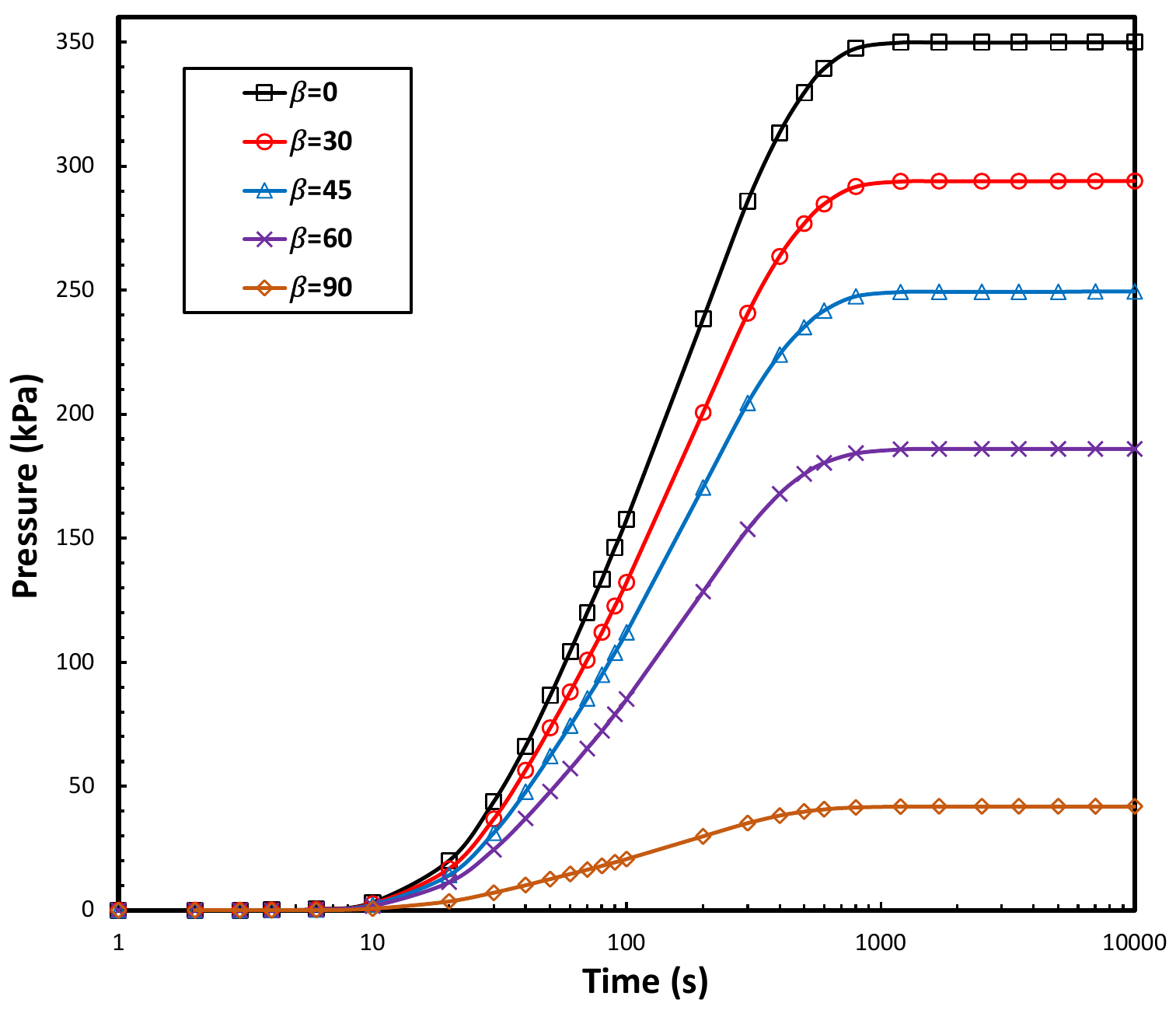}}
\caption{The pressure evolution at a) bottom and b) top of the discontinuity, and c) associated pressure difference.}
\label{fig:ex5-1pressures}
\end{figure}

The distribution contours including the vertical displacement, pressure and temperature fields for $\beta=45^{\rm{o}}$ are shown in Fig. \ref{fig:ex5-1contours}. As observed, the discontinuity associated with the fault zone is evident in all three fields, where the Heaviside enrichment is applied. Consistently, Fig \ref{fig:ex5-1-temp} shows the flow streamlines, initially aligned vertically, are severely diverted due to the presence of the impermeable fault, before retrieving their initial orientation again. In Fig. \ref{fig:ex5-1pressures}, the evolution of pore pressure at two neighboring mid-points, positioned at either side of the fault (indicated in Fig. \ref{fig:ex-5-1 geometry}), is plotted for all fault angles. As Fig. \ref{fig:ex5-1-pbot} shows, for smaller inclination angles $\beta$, the built-up pressure is greater at the bottom edge of the fault, where the fluid injection occurs. A reversed pattern is observed for pressure development on the top edge of the discontinuity (Fig. \ref{fig:ex5-1-ptop}). This mechanism is attributed to the fact that the more inclined a fault is \textendash with respect to the flow streamlines \textendash the more likely it is to hinder the fluid flow and therefore, the pressure is built up at a higher rate (Fig. \ref{fig:ex5-1-pdiff}).

\begin{figure}\centering
\subfloat[]{\label{fig:ex5-1-tbot}\includegraphics[width=.49\linewidth]{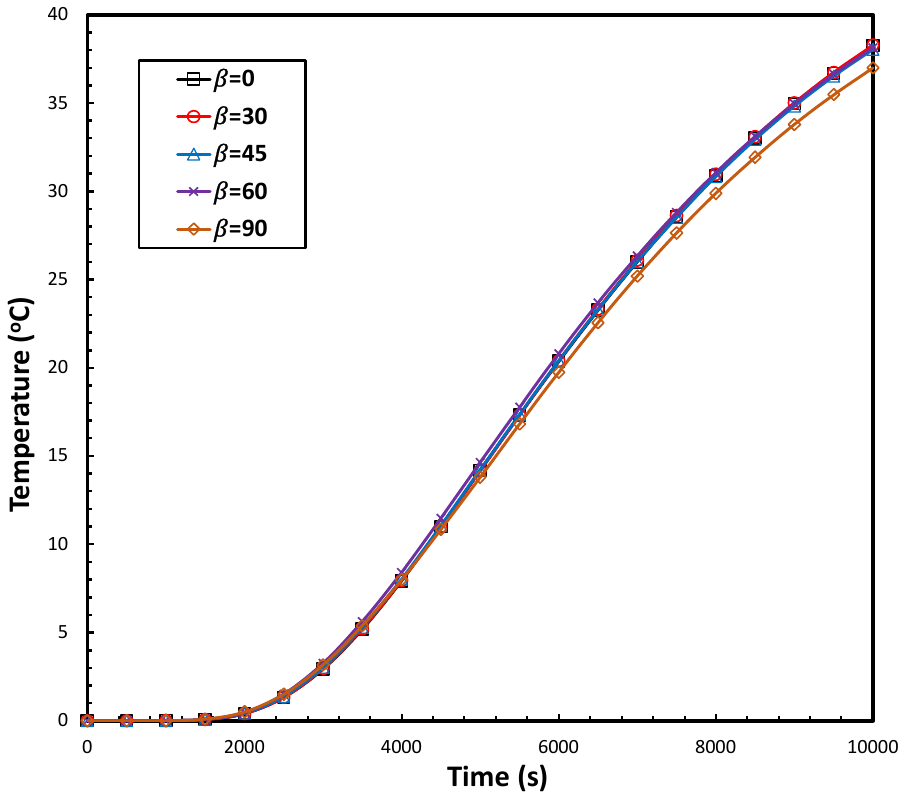}}\hfill
\subfloat[]{\label{fig:ex5-1-ttop}\includegraphics[width=.5\linewidth]{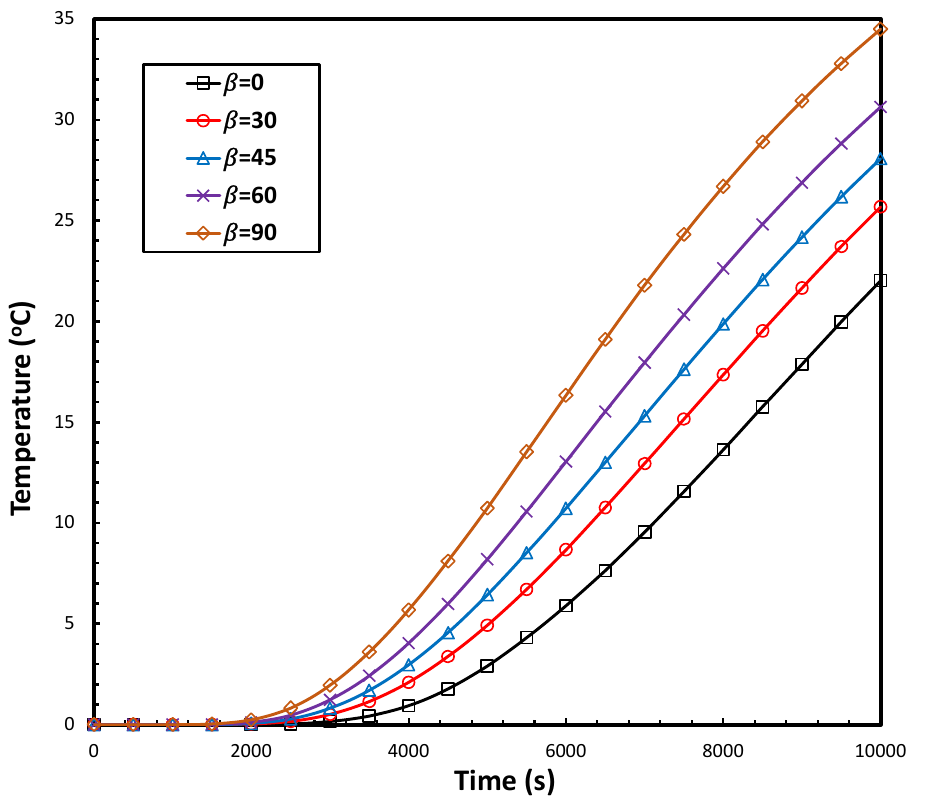}}\par 
\subfloat[]{\label{fig:ex5-1-tdiff}\includegraphics[width=.49\linewidth]{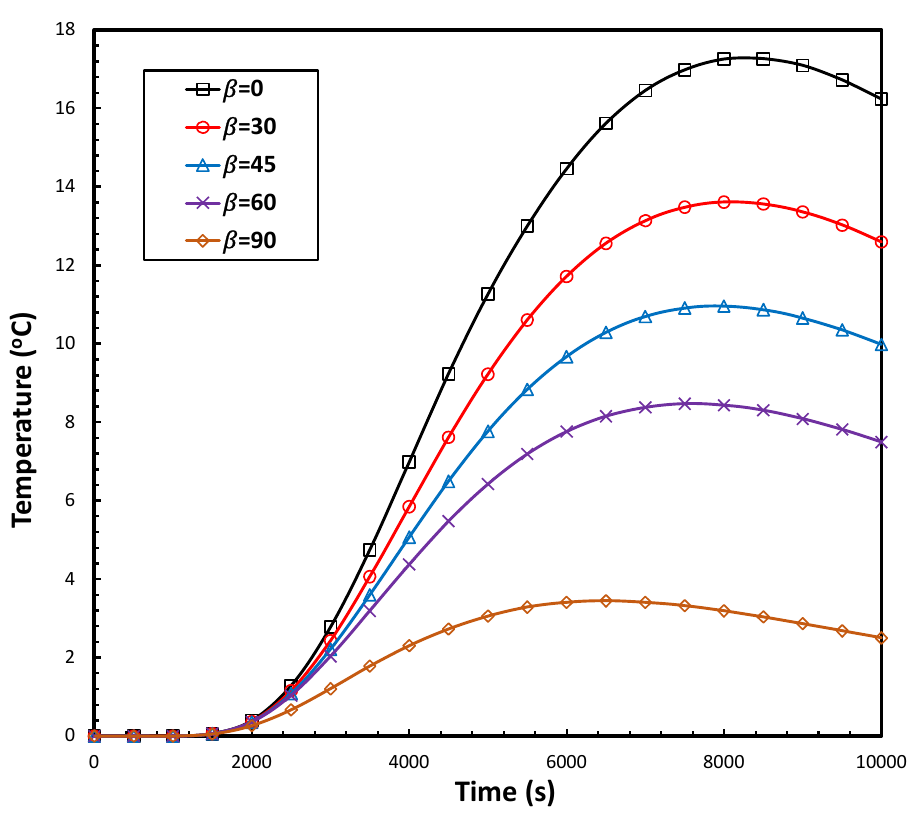}}
\caption{The temperature evolution at a) bottom and b) top of the discontinuity, and c) associated temperature difference.}
\label{fig:ex5-1temp}
\end{figure}

Fig. \ref{fig:ex5-1temp} elaborates the time-history of the temperature on either side of the fault, as well as the corresponding temperature difference. While the fault angle is observed to have a negligible effect on the temperature field along the lower face of the fault (Fig. \ref{fig:ex5-1-tbot}), the upper face undergoes significant changes in the temperature evolution (Fig. \ref{fig:ex5-1-ttop}). This can be attributed to the fact that the convective heat transfer to this zone is delayed as a result of the presence of an impermeable discontinuity that acts as a thermal barrier. In resemblance to the pressure field, as the fault inclination angle is reduced the temperature jump due to the presence of the discontinuity is decreased. As shown in Fig. \ref{fig:ex5-1-tdiff}, the greatest temperature jump is related to the case of horizontal fault. 
In all cases, the temperature jump eventually decreases to its steady-state condition, when the convective heat diffusion is fully developed throughout the entire domain.

\subsubsection{Randomly distributed faults in 2D porous media}
\label{S:4-5-2 (THM)}

\begin{figure}[!t]
\centering\includegraphics[width=0.5\linewidth]{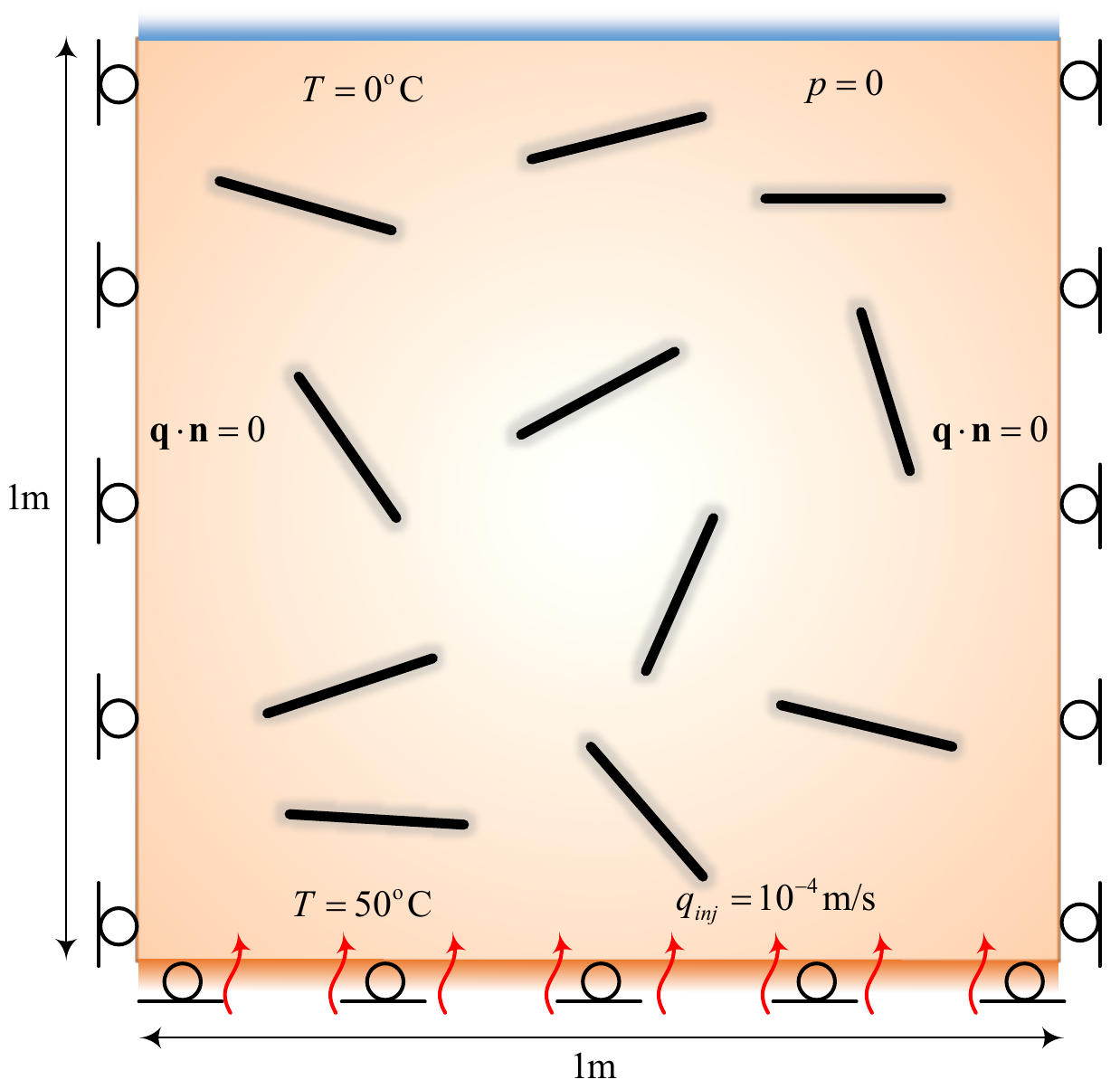}
\caption{Randomly distributed faults in a porous medium: problem geometry and boundary conditions. }
\label{fig:ex-5-2 geometry}
\end{figure}

\begin{figure}\centering
\subfloat[]{\label{fig:ex5-2-pressure}\includegraphics[width=.495\linewidth]{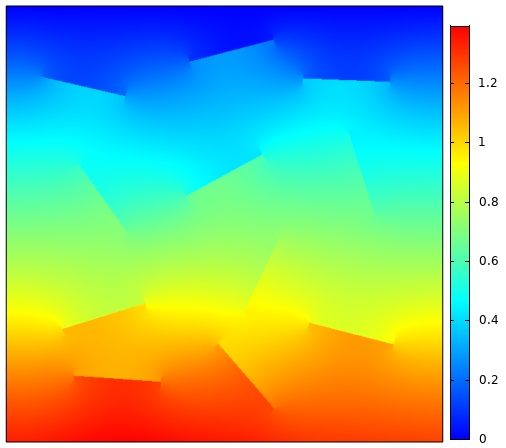}}\hfill
\subfloat[]{\label{fig:ex5-temp}\includegraphics[width=.49\linewidth]{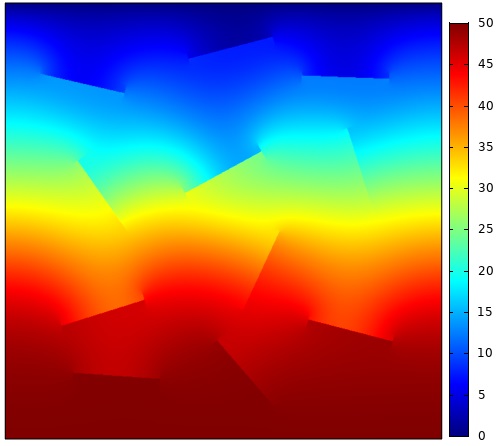}}\par 
\caption{The distribution contours of a) pressure (MPa) and b) temperature ($^ \circ {\rm{C}}$) at the end of thermo-hydro-mechanical analysis of multiple faults problem.}
\label{fig:ex5-2-contours}
\end{figure}

\begin{figure}[!t]
\centering\includegraphics[width=0.47\linewidth]{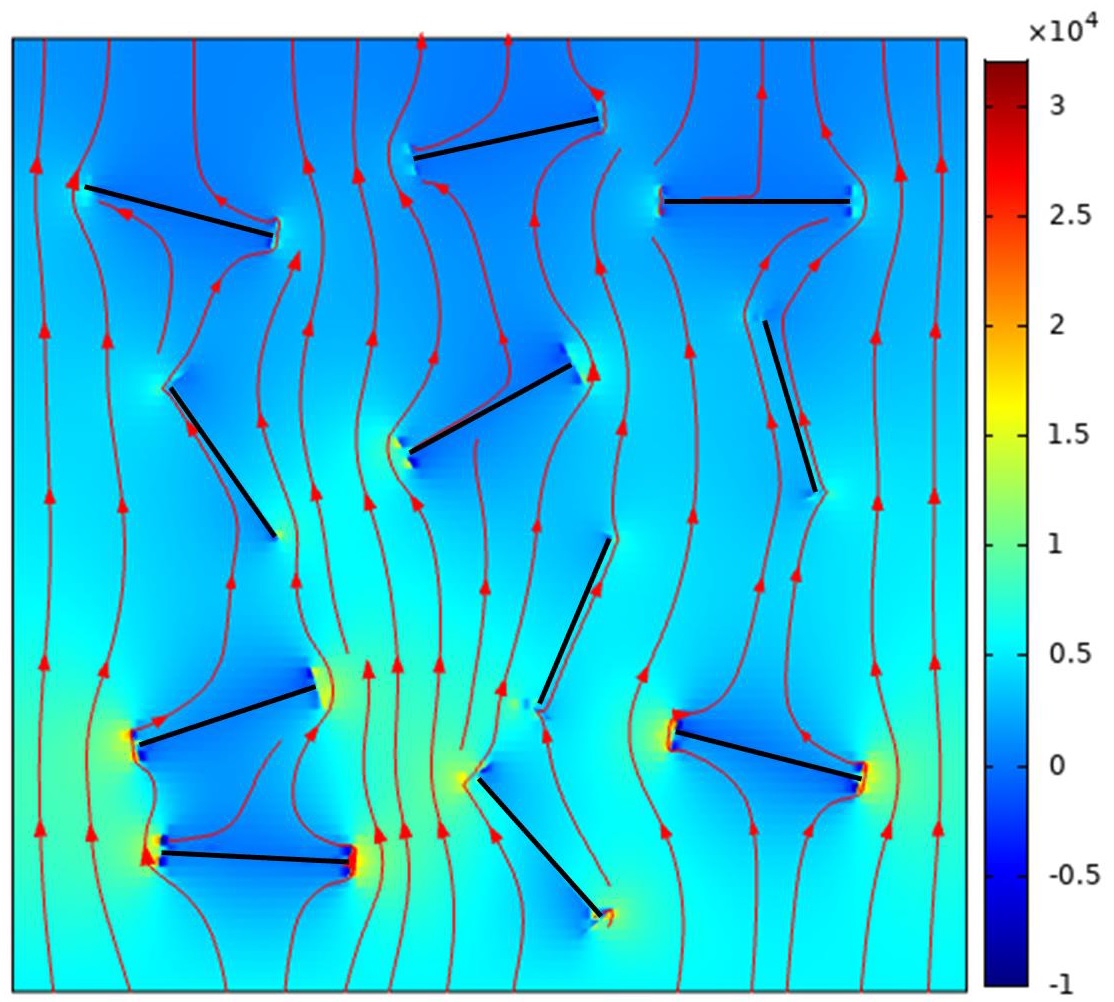}
\caption{The final distribution of heat flux contour and its associated fluid flow streamlines (values are in ${\rm{N/s}}{\rm{.m}}$).}
\label{fig:ex5-2-flux}
\end{figure}

Natural geological formations are heterogeneous and often contain numerous impermeable faults and/or inclusions (e.g., hard rocks), which could be scattered throughout their bulk. \cite{vahab2021numerical}. The following example illustrates the immense capability of the proposed implementation strategy in dealing with the simulation of multiple faults in porous media (Fig. \ref{fig:ex-5-2 geometry}). Similar boundary conditions and mesh configuration are adopted in a replica of the previous example. The domain contains 11 equally sized faults with a length of 0.2 m each, which are studied for a total simulation time of 8000 s. Fig. \ref{fig:ex5-2-contours} shows the pressure and temperature contours at the end of the simulation. The promising performance of the strong discontinuity enrichment can be easily recognized across the faults in both fields. The distribution of the heat flux accompanied by the flow streamlines is depicted in Fig. \ref{fig:ex5-2-flux} at the final stage of the analysis. Evidently, the fluid flow and the heat flux are both prevented across the contained adiabatic impermeable discontinuities, resulting in heat flux concentration at the crack tip zones.

\subsubsection{Faults in three dimensional porous media}
\label{S:4-5-3 (THM) 3D}

\begin{figure}[!t]
\centering\includegraphics[width=0.55\linewidth]{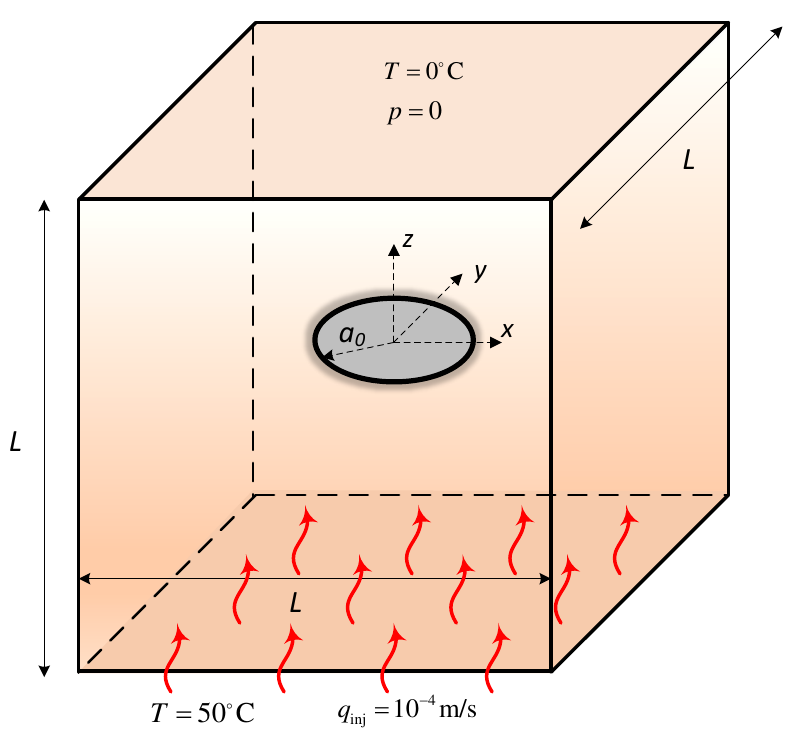}
\caption{Three dimensional thermo-hydro-mechanical simulation of a fault in porous media; problem geometry and boundary conditions. }
\label{fig:ex-5 3Dgeometry}
\end{figure}

\begin{figure}\centering
\subfloat[]{\label{fig:ex5-3Ddisp}\includegraphics[width=.5\linewidth]{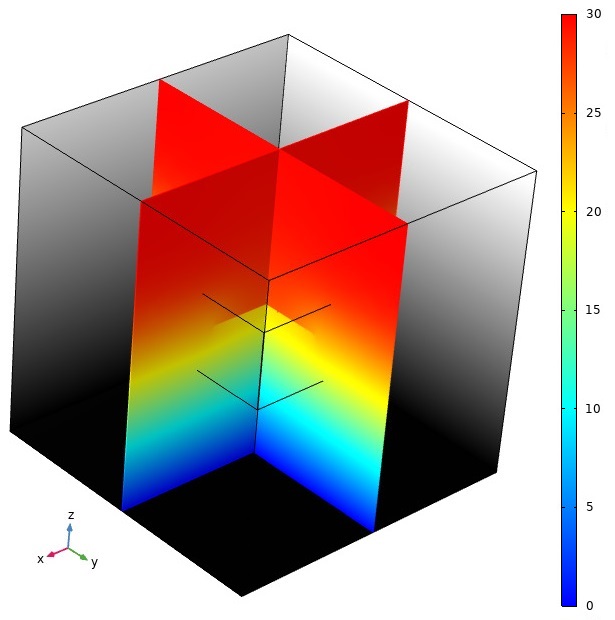}}\hfill
\subfloat[]{\label{fig:ex5-3Dpressure}\includegraphics[width=.5\linewidth]{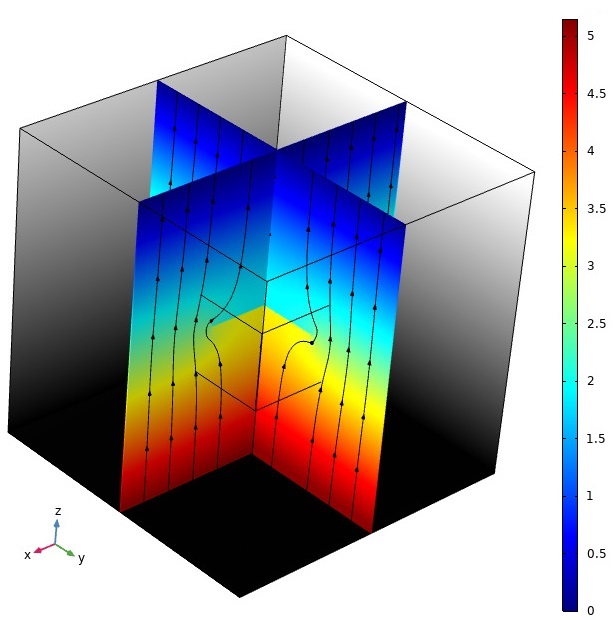}}\par 
\subfloat[]{\label{fig:ex5-3Dtemp}\includegraphics[width=.52\linewidth]{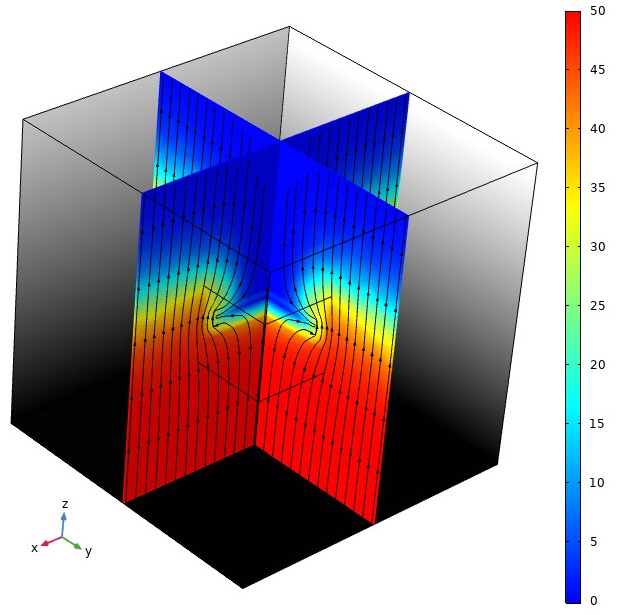}}
\caption{3D thermo-hydro-mechanical simulation results on two perpendicular planes; a) distribution of vertical displacement $u_{z}$ (mm), b) pressure distribution (MPa) together with fluid flow streamlines, and c) temperature distribution ($^ \circ {\rm{C}}$) with heat flux streamlines. }
\label{fig:ex53Dcontours}
\end{figure}

The final example is presented to show the general nature of the proposed thermo-hydro-mechanical XFEM framework to the study of three dimensional problems in geomechanics. Suppose a cubic domain with a side length of 50 m that encompasses a penny-shaped impermeable discontinuity of diameter $2{a_0} = 20$ m located at its center (Fig. \ref{fig:ex-5 3Dgeometry}). All material properties are the same as the previous examples. The bottom surface is subjected to a prescribed temperature of 50 ${^ \circ }{\rm{C}}$ and a constant inflow rate of $10^{-4}$ m/s, while both the pressure and temperature are assumed to vanish over the top surface (i.e, $p=0$; $T=0$). The remainder surfaces are considered to be undrained, with no fluid flow/heat flux. All faces of the domain are constrained in their corresponding normal directions, except the top surface, so as to emulate the in-situ boundary conditions. The porous domain is discretized using 2,312 brick elements, clustered in the vicinity of the internal discontinuity with an average element size of 1.5 m, in conjunction with 45,415 tetrahedral elements, elsewhere, that is simulated for the total duration of $3\times{10^5}$ s. Fig. \ref{fig:ex53Dcontours} illustrates the contours of vertical displacement $u_{z}$ as well as pressure and temperature fields, over two cross-shaped planes perpendicular to the discontinuity. Clearly, the discontinuity induced by the penny-shaped inclusion can be observed in all three distribution contours. This is further elaborated by noting the flow and heat flux streamlines that are depicted in Figs. \ref{fig:ex5-3Dpressure} and \ref{fig:ex5-3Dtemp}, where a diversion from the far-field vertical alignment can be observed adjacent to the discontinuity region. The promising results presented here showcase the flexibility of the implemented technique in dealing with intricate scenarios in the 2D/3D thermo-hydro-mechanical analysis of porous media with single/multiple faults.

\section{Conclusions}
\label{S:5 (Conclusions)}

In this study, an XFEM implementation for modeling thermo-hydro-mechanical problems in COMSOL Multiphysics commercial software is presented. The framework is applied for multi-field fracture analysis in 2D and 3D settings. The enrichment strategy is implemented by exploiting specific weak forms of the governing equations, which are consistent with the structure of the software. In order to take account of the standard and enriched solution fields, two sets of distinct "Solid Mechanics", "Darcy’s Law" and "Heat Transfer in Porous Media" physics interfaces associated with the mechanical deformation of the bulk, fluid flow within the pores and heat transfer through the matrix are adopted, respectively. COMSOL’s built-in features and external MATLAB functions are employed in the modeling procedures of preprocessing and level-set updating, various coupling effects between the physics and postprocessing.

Several numerical examples are presented to investigate the robustness and accuracy of the proposed implementation in geomechanical applications. In the first example, the crack growth in a bi-material composite beam subjected to thermal loading is studied and validated against the experimental observations. The second example investigates the thermo-mechanical analysis of a cracked plate, in which a comprehensive verification against an exact analytical solution is carried out. The coupled thermo-mechanical analysis is further elaborated in the third example to explore the thermo-mechanical contact within a doubled clamped beam. The next example is devoted to the hydro-mechanical modeling of discontinuities in porous media. Finally, a set of examples investigating the thermo-hydro-mechanical XFEM modeling of single/multiple discontinuities in 2D/3D deformable porous media are presented in the last example. In this fashion, the robustness and superiority of the proposed XFEM framework in dealing with intricate real-world multi-field problems is demonstrated. The simplicity of the proposed algorithms, the feasibility of inclusion of further physics, and the ease of incorporation of alternative enrichment strategies facilitate the use of state-of-the-art XFEM simulation of multiphysics, to the extent that the need for in-house codes could be alleviated. Showcasing the practicality of the inclusion of the enumerated scenarios in COMSOL Multiphysics is left to future studies.


\appendix
\section{}
\label{S:6 Appendix}

For a non-isothermal cracked domain, the stress intensity factor (SIF) can be obtained based on the J-integral concept as \cite{anderson2017fracture, duflot2008extended}
\begin{equation}
\label{eq:J1}
J = \int\limits_{{\Gamma _{\rm{J}}}} {\left[ {{w_{{\rm{mec}}}} \cdot {n_{x'}} - ({\boldsymbol{\sigma }} \cdot {\nabla _{x'}}{\bf{u}}) \cdot {\bf{n}}} \right]} d\Gamma 
\end{equation}
where ${w_{{\rm{mec}}}} = \frac{1}{2}{\boldsymbol{\sigma }}:({{\boldsymbol{\varepsilon }}}-{{\boldsymbol{\varepsilon }}_{{\rm{T}}}}){\bf{I}}$ is the \textit{mechanical} strain energy density. In addition, ${\bf{n}}$ and ${{n_{x'}}}$ are the outward normal vector and its horizontal component associated with the integral path ${{\Gamma _{\rm{J}}}}$, that encloses the crack tip (in the local crack coordinates $x’-y’$). For mixed-mode loading in two dimensions, the relation between the stress intensity factors and J-integral can be expressed as,
\begin{equation}
\label{eq:J2}
J=\frac{K_{\rm{I}}^{2}+K_{\rm{II}}^{2}}{{E}'}
\end{equation}
where ${E}'$ is defined as $E/(1-\upsilon ^2)$ and $E$ for plane strain and plane stress problems, respectively.

The stress intensity factors can be calculated by means of the interaction integral method with satisfactory accuracy \cite{anderson2017fracture}. Accordingly, an auxiliary displacement field is employed in addition to the actual displacement field of the problem, denoted by states (2) and (1), respectively, such that
\begin{equation}
\label{eq:I12}
I^{(1+2)}=\int_{\Gamma _{\rm{J}}}^{}\left [ W^{(1,2)}\cdot n_{{x}'}-(\boldsymbol{\sigma }^{{(1)}}\nabla_{{x}'}\mathbf{u}^{(2)}+\boldsymbol{\sigma }^{{(2)}}\nabla_{{x}'}\mathbf{u}^{(1)})  \right ]d\Gamma 
\end{equation}
where $W^{(1,2)} = \boldsymbol{\sigma }^{(1)}\cdot \boldsymbol{\varepsilon }^{(2)}=\boldsymbol{\sigma }^{(2)}\cdot \boldsymbol{\varepsilon}^{(1)}$ denotes the interaction strain energy. Combining Eqs. \ref{eq:J1}, \ref{eq:J2} and \ref{eq:I12}, the interaction integral is related to the SIFs corresponding to the actual and auxiliary states as
\begin{equation}
\label{eq:I12K}
I^{(1+2)}=\frac{2}{{E}'}(K_{\text{I}}^{(1)}K_{\text{I}}^{(2)}+ K_{\text{II}}^{(1)}K_{\text{II}}^{(2)})
\end{equation}
The stress intensity factors of a mixed-mode crack domain can be therefore calculated by taking advantage of the pure mode I (i.e., $K_{\rm{I}}^{(2)}=1$, $K_{\rm{II}}^{(2)}=0$) or pure mode II (i.e., $K_{\rm{I}}^{(2)}=0$, $K_{\rm{II}}^{(2)}=1$) asymptotic solutions. Moreover, the alternative domain form of Eq. \ref{eq:I12} can also be employed by the use of special weighting functions \cite{khoei2012thermo}.

In this study, the internal variables and built-in mathematical functions of COMSOL are employed to carry out the stress intensity factor calculations. To this end, a list of all required variables is defined, which includes: normal vector components, actual and auxiliary stress and strain fields and associate derivatives, the displacement field, the crack tip coordinates, the interaction strain energy density, etc. In this study, the numerical integration of Eq. \ref{eq:I12} is carried out by exploiting the built-in circular path integration operator \textit{circint} in COMSOL. Alternatively, the equivalent domain form of interaction integral can be applied using the \textit{diskint} operator for a circular area encompassing the crack tip.
\bibliographystyle{model1-num-names}
\bibliography{references.bib}

\end{document}